\def\qut#1{``#1''}

\documentclass{copernicus} 
\usepackage{graphicx,lineno,url}
\usepackage{color,ulem,bm} 

\begin{document}

\runninglinenumbers 

\titleheight{10.0cm}

\title{Recent variability of the solar spectral irradiance and its impact on climate modelling}

\author[1]{I.~Ermolli}
\author[2]{K.~Matthes}
\author[3]{T.~Dudok~de~Wit}
\author[4]{N.~A.~Krivova}
\author[5]{K.~Tourpali}
\author[6]{M.~Weber}
\author[7]{Y.~C.~Unruh}
\author[8]{L.~Gray}
\author[9]{U.~Langematz}
\author[10]{P.~Pilewskie}
\author[11,12]{E.~Rozanov}
\author[11]{W.~Schmutz}
\author[11]{A.~Shapiro}
\author[4,13]{S.~K.~Solanki}
\author[10]{T.~N.~Woods}

\affil[1]{INAF, Osservatorio Astronomico di Roma, Monte Porzio Catone, Italy}
\affil[2]{GEOMAR I Helmholtz-Zentrum f\"ur Ozeanforschung Kiel, Kiel, Germany}
\affil[3]{LPC2E, CNRS and University of Orl\'eans, Orl\'eans, France}
\affil[4]{Max-Planck-Institut f\"ur Sonnensystemforschung, 37191 Katlenburg-Lindau, Germany}
\affil[5]{Laboratory of Atmospheric Physics, Aristotle University of Thessaloniki, Greece}
\affil[6]{Institut f\"ur Umweltphysik, Universit\"at Bremen FB1, Bremen, Germany}
\affil[7]{Astrophysics Group, Blackett Laboratory, Imperial College London, SW7 2AZ, UK}
\affil[8]{Centre for Atmospheric Sciences, Dept. of Atmospheric, Oceanic and Planetary Physics, University of Oxford, UK}
\affil[9]{Institut f\"ur Meteorologie, Freie Universit\"at Berlin, Berlin, Germany}
\affil[10]{University of Colorado, Laboratory for Atmospheric and Space Physics, Boulder, CO, USA}
\affil[11]{Physikalisch-Meteorologisches Observatorium, World Radiation Center, Davos Dorf, Switzerland}
\affil[12]{IAC ETH, Zurich, Switzerland}
\affil[13]{School of Space Research, Kyung Hee University, Yongin, Gyeonggi 46-701, Republic of Korea}

\correspondence{I.~Ermolli (ilaria.ermolli@oa-roma.inaf.it)}

\runningtitle{Spectral irradiance and climate}

\runningauthor{I.~Ermolli et~al.}

\received{29 July 2012}
\accepted{13 March 2013}
\published{}

\firstpage{1}

\maketitle

\abstract  
      The lack of long and reliable time series of solar spectral irradiance (SSI) measurements makes an accurate quantification of solar contributions to recent climate change difficult. Whereas earlier SSI observations and models  provided a qualitatively consistent picture of the SSI variability, 
recent measurements by  the SORCE satellite suggest  a significantly stronger variability in the ultraviolet (UV) spectral range  and changes in the visible and near-infrared (NIR) bands in anti-phase with the solar cycle. A  number of recent chemistry-climate model (CCM) simulations have shown  that this might have significant implications on the Earth's  atmosphere. Motivated by these results, we summarize here our current knowledge of SSI variability and its
      impact on Earth's climate. 
      
We present a detailed overview of existing SSI measurements and provide thorough comparison of models available to date.  SSI changes influence the Earth's atmosphere, both directly, through changes in shortwave (SW) heating and therefore, temperature and ozone distributions in the stratosphere, and indirectly, through  dynamical feedbacks. We investigate these direct and indirect effects using several state-of-the art CCM simulations forced with  measured and modeled SSI changes.   A unique asset of this study is the use of a common comprehensive approach for an issue that is usually addressed separately by different communities. 

We show  that the SORCE measurements are  difficult to reconcile with earlier observations  and  with SSI models. 
 Of the five
SSI models discussed here, specifically NRLSSI, SATIRE-S, COSI, SRPM, and OAR, only one  shows a behaviour of the UV and visible irradiance qualitatively resembling that of the recent SORCE measurements. However, the  integral of the SSI computed with this model over the entire spectral range  does not reproduce the measured cyclical changes of the total solar irradiance, which is an essential requisite for realistic  evaluations of solar effects on the Earth's climate in CCMs. 

 We show that within  the range provided  by the recent SSI observations and semi-empirical models discussed here,  the NRLSSI model and SORCE observations represent the lower and upper limits in the magnitude of the SSI solar cycle  variation. 
 
The results of the CCM simulations, forced with the SSI solar cycle variations
estimated  from the NRLSSI model and from SORCE
measurements,  show that the direct solar response in the stratosphere is larger for the SORCE than for the NRLSSI data. Correspondingly, larger UV forcing also leads to a larger surface response.

Finally, we discuss the reliability of the available
 data  and we propose additional coordinated work,  first to build composite SSI datasets out of scattered observations and to refine current SSI models, and second, to run  coordinated CCM experiments.
 

\introduction
\label{sec:intro}
\begin{figure}
\includegraphics[width=85mm]{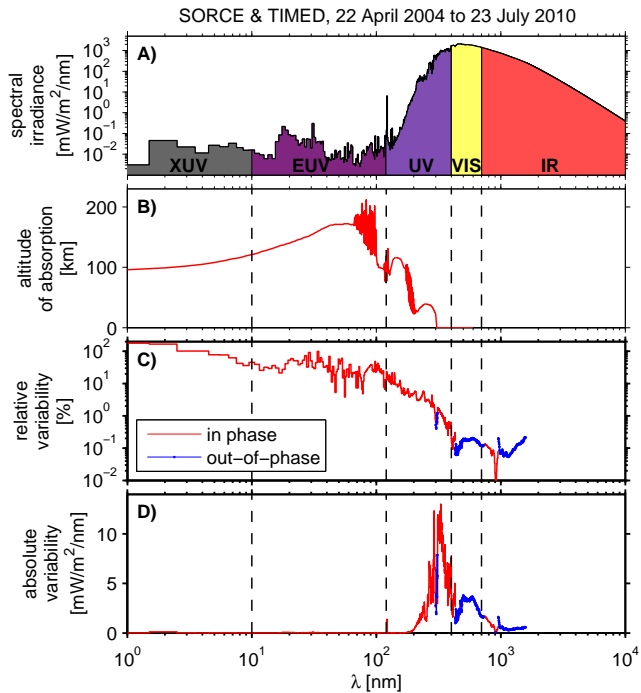}
\caption{The solar spectral irradiance as inferred from SORCE and TIMED
observations only, from 22~April 2004 till 23~July 2010.  \textbf{(A)}~shows the average
solar spectral irradiance for that period.  A~black-body model has been used
to extend the SSI for wavelengths beyond 1580\,\unit{nm}.  \textbf{(B)}~displays the
characteristic altitude of absorption in the Earth's atmosphere for each
wavelength, defined as the altitude at which the optical depth equals one.  \textbf{(C)}~shows the relative variability (peak to peak/average) for solar cycle
variations inferred from measurements obtained between 22~April 2004 and 23~July 2010. Spectral regions, where the variability is in phase with the solar
cycle (represented by, e.g. the sunspot number or the TSI) are marked in red,
while blue denotes ranges where the variability measured by SORCE is
out-of-phase with the solar cycle.  These phases, as well as the magnitude of
the variability in the UV, are not all reproduced by models and other
observations (see Sect.~\ref{sec:models} as well as
Figs.~\ref{fig:ssicontri},~\ref{fig:varia} and~\ref{fig:refrange}), and thus
should be considered with care.  \textbf{(D)}~shows the absolute variability, which
peaks strongly in the near-UV.}
\label{fig:spectrum}
\end{figure}

\begin{figure*}
\includegraphics[width=150mm]{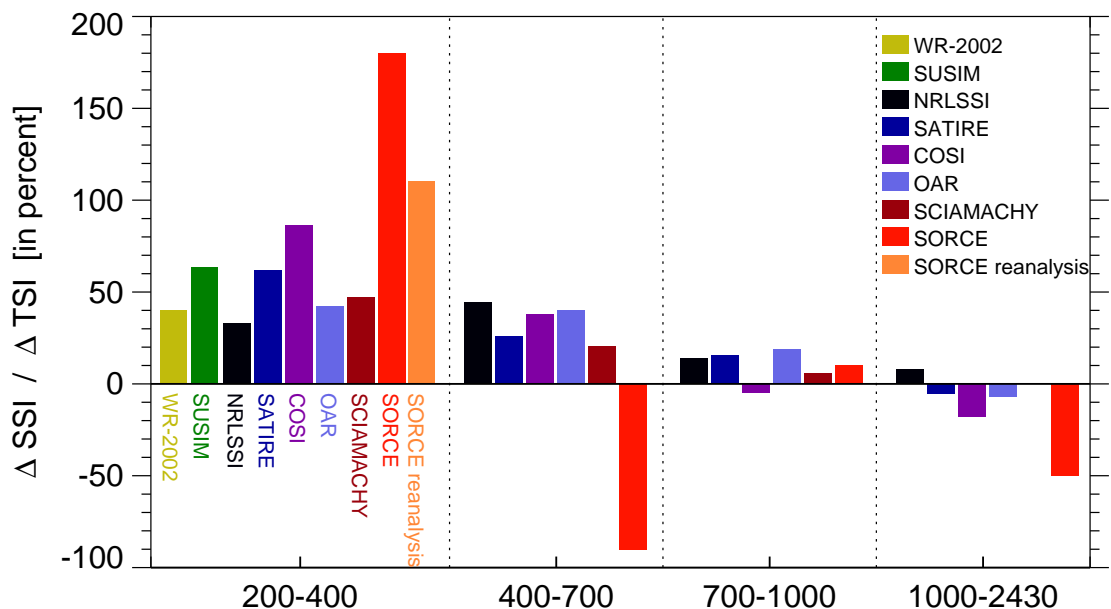}
\caption{Relative
contribution of the UV (200--400\,\unit{nm}), visible (400--700\,\unit{nm}),
near-IR (700--1000\,\unit{nm}) and IR (1000--2430\,\unit{nm}) ranges to the
TSI change over the   solar cycle  as derived from measurements and models
described in Sects.~\ref{sec:meas} and~\ref{sec:models}.  For SORCE/SIM, only
the period between 2004 and 2008 can be considered.  For other data and
models, the plotted relative differences are between solar maximum an minimum.
Within each wavelength bin, from left to right: WR-2002 \citep[light green,
only in UV;][]{WoodsRottman:2002}, UARS/SUSIM \citep[green, only in
UV;][]{Morrilletal:2011}, NRLSSI (black), SATIRE-S (blue), COSI (purple), OAR
(light blue), SCIAMACHY (brown), SORCE (red), and SORCE re-analysis
\citep[orange, only in UV;][]{Woods:2012}. The exact wavelength ranges used
for SUSIM and SCIAMACHY in the UV are 150--400\,\unit{nm} and
240--400\,\unit{nm}, respectively.  The possible related corrections are,
however, expected to lie within 2--3\,\unit{\%}.  Note that for the
SCIAMACHY-based model, the original values listed by \citet{Pagaranetal:2009}
are shown.  As discussed by \citet{KrivovaSolanki:2012}, these values should
most likely be corrected by a~factor of roughly 1.2.}
\label{fig:ssicontri}
\end{figure*}

      The question of whether -- and to what extent -- the Earth's climate
      is influenced by solar variability remains central to the
      understanding of anthropogenic climate change.   According to the 
      4th assessment report of the Intergovernmental Panel on Climate Change, 
      solar variability represents about 8\% of recent total net radiative 
      anthropogenic forcing \citep[][]{Solomonetal:2007}.  
      However, there is
      a~large uncertainty in this figure because several aspects of solar
      forcing and the different mechanisms by which solar variability
      influences the Earth's environment are still poorly understood
      \citep[see e.g.][and references
      therein]{Papetal:2004,Calisesietal:2007,Haigh:2007,Grayetal:2010,Lockwood:2012}. For
      these reasons, the quantification of  solar contribution to climate
      change remains incomplete.  This is further highlighted by some of the
      most recent investigations of solar spectral irradiance (SSI)
      variations and estimates of their influence on the Earth's atmosphere
      based on chemistry-climate model (CCM) simulations.

      Regular space-based measurements of the solar irradiance started in
      1978. 
      The 
      total solar irradiance (TSI), 
      i.e. the spectrally integrated radiative power
      density of the Sun incident at the top of Earth's atmosphere, has been
      monitored almost continuously and was found to vary on different time
      scales \citep[][]{Willsonetal:1981,FroehlichLean:2004,Froehlich:2009}.
      Most noticeable  is the $\approx$\,0.1\,{\%} modulation of TSI in phase
      with the 11-yr solar  cycle.  Changes 2-3 times larger than this  are observed on time scales shorter than few days. 
      Measurements of
      SSI, however, are not continuous over the satellite era and until
      recently have concentrated on the ultraviolet (UV) radiation, because
      of the larger relative variability of SSI below 400\,\unit{nm}         (Fig.~\ref{fig:spectrum}) and the
      impact of these wavelengths on the terrestrial atmosphere through
      radiative heating and ozone photochemistry.

      SSI variations differ from those observed in the TSI. The variability
      of visible and NIR bands barely exceeds 0.5\,{\%} over a~solar cycle;
      in the near UV and shorter wavelengths variability increases with
      decreasing wavelength, reaching several percent at
      200--250\,\unit{nm}, and several tens of percent, and even more, below
      about 200\,\unit{nm} \citep[e.g.][ and references
      therein]{Floydetal:2003a}. These bands are almost completely absorbed in the Earth's middle and upper atmosphere (Fig.~\ref{fig:spectrum}) and are the primary agent affecting 
      heating, photochemistry, and therefore, the dynamics of the
      Earth's  atmosphere.   Variations of the solar UV radiation between 120 and 350\,\unit{nm}  lead to changes in
      stratospheric ozone and heating  that amplify the effect of the UV radiation in the Earth's  atmosphere, possibly also through indirect mechanisms \citep[e.g. the  \qut{top-down}
      mechanism,][]{Grayetal:2010}. 
     Hence,  although the UV radiation shortward of 400\,\unit{nm} 
      represents less than 8\,{\%} of the TSI, its           
      variability may have a~significant impact on climate.  In contrast, 
      the visible and IR bands, which have the largest contribution to the TSI,  small variations  over the solar cycle, and  no absorption  in the atmosphere but in some well-defined IR bands, 
      directly heat the Earth's surface and the lower atmosphere.  The large amount of solar flux at the visible and IR  bands implies that small flux differences may induce important terrestrial consequences. The impact of the variability of   these bands  on
      the Earth's climate is expected to be small,  although it may involve
     amplification mechanisms  \citep[e.g. the \qut{bottom-up}
      mechanism,][]{vanLoonetal:2007,Meehletal:2009}.  

      Early satellite measurements of the solar UV variability have shown
      a~qualitatively consistent behaviour \citep{DelandCebula:2012}, which
      is fairly well reproduced by SSI models
      \citep[e.g.][]{Leanetal:1997,Krivovaetal:2006a,
      Unruhetal:2012,LeanDeland:2012}.  This
      situation changed with the launch of the Spectral Irradiance Monitor
      instrument \citep[SIM,][]{Harderetal:2005a} onboard the Solar
      Radiation and Climate Experiment satellite
      \citep[SORCE,][]{Rottman:2005} in 2003, which was shortly after the
      most recent maximum of solar activity.  The SORCE/SIM data showed
      a~four to six times greater decrease of the UV radiation between 200
      and 400\,\unit{nm} over the period 2004--2008 \citep{Harderetal:2009},
      part of the declining phase of solar cycle 23, compared to earlier
      measurements and models
      \citep[][]{Balletal:2011,Pagaranetal:2011b,Unruhetal:2012,DelandCebula:2012,LeanDeland:2012}. This
      larger decrease measured in the UV (Fig.~\ref{fig:ssicontri}), which
      exceeds the TSI decrease over the same period by almost a~factor of
      two, is compensated by an increase in the visible and NIR bands.
         Variability out-of-phase with solar activity is indeed predicted by
      some SSI models in the NIR, but with a~significantly lower magnitude than
      found by SORCE/SIM. Details are provided in the following. The inverse variability observed by SIM in a wide integrated band in the visible was, however, unexpected. It can be interpreted as a result of effects induced by the evolution of surface magnetism in the solar atmosphere \citep[e.g.][]{Harderetal:2009}.  However, other observations and analyses of existing long-term SSI data show  results in contrast with those derived from SORCE/SIM \citep[][]{Wehrlietal:2012}.

      When used as solar input to CCM simulations, SORCE/SIM observations
      lead to significantly larger shortwave (SW) heating rates in the upper
      stratosphere compared to results obtained by using the commonly
      utilised NRLSSI model data \citep[][]{Leanetal:1997,Lean:2000}, and a~decrease of stratospheric
      ozone above an altitude of 45\,\unit{km} during solar maximum
      \citep[][]{Haighetal:2010}. These changes in radiative heating and
      ozone photochemistry in the stratosphere also impact the responses of
      the \qut{top-down} solar UV mechanism in the Earth's atmosphere and at
      the surface \citep[][]{KoderaKuroda:2002}, which may depart from
      current understanding
      \citep[][]{Cahalanetal:2010,Inesonetal:2011,Merkeletal:2011,Oberlaenderetal:2012,Swartzetal:2012}. Validation of the results of these model simulations with
      ozone, zonal wind, or temperature measurements is difficult, because the data are sparse  and do not cover enough solar cycles.   Although some ozone
      observations seem to agree with model calculations
      \citep[e.g.][]{Haighetal:2010,Merkeletal:2011}, it should be noted
      that the SIM measurements  employed for the analysis  covered less than one solar cycle and
      required extrapolation over a~full cycle, and therefore, added
      uncertainty \citep[][]{Garcia:2010}. Also the transition altitude from
      in-phase (lower and middle stratosphere) to out-of-phase (upper
      stratosphere) ozone signals with the solar cycle is not consistent
      among the different models and requires further investigations.

      Unfortunately, observations of the full solar spectrum will likely
      have a~multi-year gap before the next generation SSI instrument is
      launched.  Based on data presently available, a~thorough understanding
      of the impact of SSI on climate requires verification and validation
      of existing SSI measurements for internal consistency, calculations of
      middle atmosphere climate models with different reliable scenarios of
      SSI variations, and comparison of measurements and model results with
      climate records, i.e.  a~study involving the coordinated work of
      various research communities, which is part of the COST
      Action ES1005\footnote{TOSCA -- towards a~more complete assessment of
      the impact of solar variability on the Earth's climate,
      \url{http://www.tosca-cost.eu}.} to which most of the authors of this
      paper belong.

      This paper  summarises and compares, for the
      first time, a~large number of SSI observations and models, and
      discusses the impact of these data on Earth's climate.   A review in this area was recently published by  \citet[][]{Grayetal:2010}, but their focus was more on  different possible forcing mechanisms.  Although the Sun affects the climate system in numerous ways, we  here focus   on
      radiative forcing only, with particular attention given to the role of
      the SSI rather than that of the TSI, which is still the sole solar
      input in many climate models.   Since space-based observations of the SSI and
      of the terrestrial atmosphere conditions are sparse or absent before
      1980,  we restrict our analysis to the data of the last three
      decades, i.e.  roughly the last three solar cycles.  Furthermore, we
      concentrate on the effects of the UV variability because of its
      potentially large impact on the terrestrial atmosphere. 

      The paper is organised as follows: in Sect.~2, measurements of SSI
      variations are described, and their accuracy on time scales from days
      to 11-yr solar cycle is discussed.  In Sect.~3 we delineate
      mechanisms responsible for the SSI variations, outline methods of
      irradiance reconstructions and briefly describe and compare several of
      the most broadly used models of SSI variations. Section~4 discusses
      the impact of the current lower and upper boundaries of SSI solar
      cycle estimated variations on the atmospheric response in several current state-of-the art CCMs.  A~summary and
      concluding remarks are provided in Sect.~5.

\section{Solar irradiance measurements}
\label{sec:meas}

      We first discuss the available solar irradiance measurements, focussing on the most recent data. These data are input to and, at the same time, the main source of constraints for both the SSI models and CCM simulations presented below. 
We describe
      the evolution of measurements carried out from space since 1978
      and discuss major instrumental and
      measurement problems that limit the creation of single composite time
      series from existing records  (Sect.~\ref{sec.timeseries}). 
       Long and accurate SSI  time series are critical to obtain reliable  estimates of the solar variability impact on the terrestrial atmosphere. 
      We then present instruments and
      techniques used to derive four SSI data sets that are employed for CCM
      simulations    (Sect.~\ref{sec.scia}). Finally, we     discuss recent re-assessment of the TSI absolute value (Sect.~\ref{sec:tsi},) since it enters some climate and SSI models,  and outline the perspectives  of future  SSI observations (Sect.~\ref{sec.24}).

 \subsection{Observation of solar irradiance}
\label{sec.timeseries}
\begin{figure*}
\includegraphics[width=168mm]{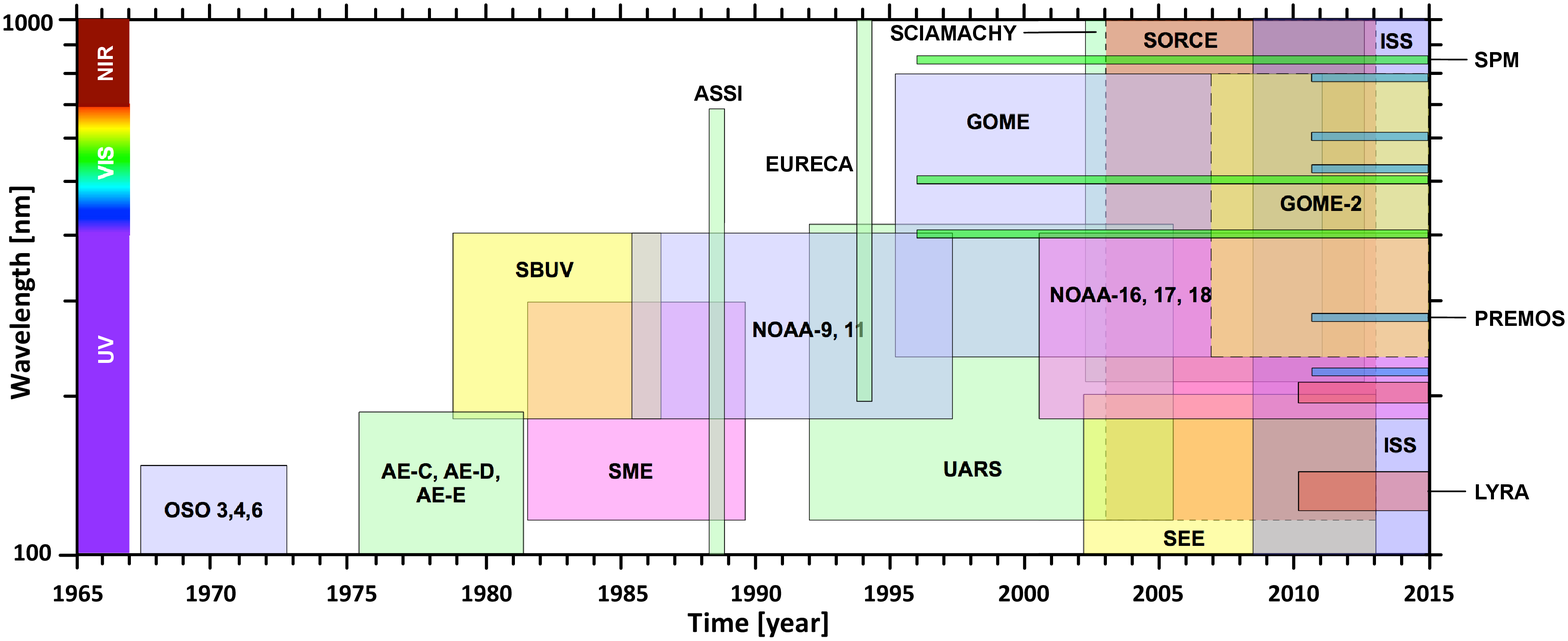}
\caption{Overview of the main
satellite missions that have made SSI observations at wavelengths higher than
100\,\unit{nm}. Short missions such as sounding rockets are not indicated.}
\label{fig:ssioverview}
\end{figure*}

      For many centuries, the Sun has been considered as an example of
      stability and, not surprisingly, the TSI, i.e. the quantity of
      radiative power density (\unit{W\,m^{-2}}) at normal incidence on top
      of the atmosphere, at a~Sun--Earth distance of one astronomical unit,
      has been called  solar constant. This term is still often used. However,
      regular space-based observations that started in 1978
      \citep[][]{Willsonetal:1981} have revealed that the TSI varies over
      time scales of minutes to decades, and probably even longer
      \citep[e.g.][]{Froehlich:2006,Froehlich:2009}.

      The most  noticeable  variation of the TSI is a~0.1\,{\%} modulation in
      phase with the solar activity or sunspot cycle. Although quantifying
      such small variations is a~major technological  challenge, it is
      strongly motivated by the desire to understand solar variability, and
      even more so by the importance of the TSI for terrestrial energy
      budget studies \citep[][]{Trenberth:2009}.

      The TSI is the spectral integral of SSI over all wavelengths but its
      weak variability masks the fact that relative SSI variations show
      a~strong wavelength dependence (Fig.~\ref{fig:spectrum}). In
      particular, the visible and NIR bands are the least variable of the
      solar spectrum with a~relative solar cycle amplitude of the same order
      as for the TSI (0.1\,{\%}), whereas values of 1 to  100\,{\%} are
      observed in the UV variations, and in excess of 100\,{\%} in the soft
      X-ray range (below 10\,\unit{nm}). Each individual spectral band has
      a~markedly different impact on the terrestrial atmosphere, which 
      depends on the atmospheric processes affected by the given band, the  amount  of the spectral flux, 
      and its variation.

 Figure~\ref{fig:ssioverview} shows an overview of all SSI space experiments
    as a~function of the period of observations
      and wavelength coverage (above 100\,\unit{nm}).      Regular space-based monitoring of the solar spectrum over a~broad
      range, covering, in addition to the UV, the visible and the IR up to
      2.4\,\unit{\mu m}, started with the launch of ENVISAT/SCIAMACHY in 2002 and
      SORCE/SIM in 2003. Measurements in the UV below 400\,\unit{nm} began
      several decades earlier. Most of these
      earlier measurements, however, are difficult to use for quantifying
      the solar variability, for reasons that will be detailed below.

      The longest records of SSI measurements were provided by SOLSTICE and
      SUSIM aboard the UARS (Upper Atmosphere Research Satellite) spacecraft
      \citep[][]{Rottmanetal:2004}.  These instruments observed solar UV
      radiation between 120 and 400\,\unit{nm} from 1991 to 2001 and 2005,
      respectively.  These measurements pointed to the importance of the
      irradiance variations in the UV
      \citep[][]{Woodsetal:2000a,Rottmanetal:2001,Floydetal:2002a,Floydetal:2003a},
      although the solar cycle variability of solar radiation above
      approximately 250\,\unit{nm} remained relatively uncertain due to
      insufficient long-term stability of the instruments
      \citep[][]{Woodsetal:1996}. In addition to SOLSTICE and SUSIM, there
      is the long time series of 200--400\,\unit{nm} solar UV observations
      by several NOAA SBUV instruments, which had { underflight} calibrations
      aboard the Space Shuttle \citep[][]{Cebulaetal:1998}.  Table 1  of \citet[][]{DelandCebula:2012} summarises the measurement uncertainties for these instruments.

      Regular observations of the visible and NIR bands covering more than
      one year started with SOHO/VIRGO-SPM \citep[][]{Froehlichetal:1997} in
      1996 at three selected bands and continued with the ERS/GOME
      \citep[][]{Weberetal:1998, Burrowsetal:1999}, ENVISAT/SCIAMACHY
      \citep[][]{Bovensmannetal:1999}, SORCE/SIM
      \citep[][]{Harderetal:2005a,Harderetal:2005b}, and ISS/SOLSPEC
      \citep[][]{Thuillieretal:2009} instruments.  These SSI missions are described in more
      detail   below.

      Merging all UV observations into a~single homogeneous composite record
      is a~major challenge \citep[][]{Delandetal:2004} that is hampered by
      several problems. First, the lifetime of most instruments does not
      exceed a~decade. This makes long-term observations covering periods
      that exceed single instrument lifetimes, of prime interest for climate
      models, very difficult. A~second obstacle is the differing
      technologies and modes of operation of various space-based
      instruments. The cross-calibration of individual records is further
      hampered by the fact that overlapping observations disagree, and the existing data records are spectrally
      and temporally intermittent. Although missing observations can be
      filled in by using data regression based on time series of solar
      proxies such as the Mg~II index, which are well correlated with UV
      variations \citep[][]{DelandCebula:1993,Vierecketal:2001,Lean:1997b},
      none of the existing solar proxies can properly reproduce solar
      irradiance in a~spectral band on all time scales
      \citep[][]{Dudokdewitetal:2009}.

      The most critical issue for all SSI instruments is the optical
      degradation caused by the energetic radiation in the space
      environment. Two options have been employed to account for
      instrumental degradations.  The first one is to provide redundancy in
      the instrument design, by using, e.g. a~dual spectrometer setup with
      detectors that experience different accumulated exposure time or by
      planning redundancies in spectral channels and calibration lamps from
      which degradation corrections can be derived.  This approach was used,
      for instance, for UARS/SUSIM and SORCE/SIM
      \citep[][]{Brueckneretal:1993,Harderetal:2005a}.  The second option is
      to use { reference data, e.g. solar irradiance measurements for long-term calibrations, as reported for the NOAA SBUV instruments,  periodic recalibrations using sounding rockets, or} stable external calibration targets like selected stars, as  
      done for UARS/SOLSTICE and SORCE/SOLSTICE
      \citep[][]{Mcclintocketal:2005}.  Sensitivity changes and degradation
      are strongly wavelength-dependent, which makes creating a~properly
      cross-calibrated SSI record very difficult.  One attempt for
      a~composite SSI time series in the UV is provided by
      \citet[][]{DelandCebula:2008}.

      Cross-calibration of different SSI records is also limited by the lack
      of realistic confidence intervals for the existing data. This aspect
      has been thoroughly investigated for the TSI   
      and sound estimates have been obtained for time scales from days to
      years. No meaningful estimates, however, exist for time scales
      exceeding one decade, which are needed for climate studies. The
      situation is much worse for SSI measurements, whose relative
      uncertainties often are two to three orders of magnitude larger.

      In the following subsections a~brief summary of {  newly available  SSI
      observations obtained during the last decade} is given. Also a~brief description of recent developments
      regarding the absolute values of the TSI, and its variability is
      provided.

\subsection{SSI time series}
\label{sec.scia}

\subsubsection{SCIAMACHY and GOME}

      SCIAMACHY (SCanning Imaging Absorption spectroMeter for Atmospheric
      CHartographY) and GOME (Global Ozone Monitoring Experiment) onboard
      the ENVISAT and ERS-2 satellites, respectively, are atmospheric
      sounders that measure terrestrial atmospheric trace gases
      \citep[][]{Burrowsetal:1999, Bovensmannetal:1999,
      Bovensmannetal:2011}. The primary purpose of direct solar measurements
      by SCIAMACHY and the two GOMEs (a~second GOME is flying  on
      METOP-A since 2006 and a third GOME on METOP-B since 2012) is to Sun-normalise the backscattered light from the
      terrestrial atmosphere, which is then inverted to determine
      atmospheric trace gas amounts. This normalisation does not require
      absolute radiometric calibration and cancels out degradation
      effects. The SCIAMACHY and GOME instruments have been radiometrically
      calibrated before launch.

      In order to provide estimates for solar cycle variability from
      SCIAMACHY measurements (230\,\unit{nm}--2.4\,\unit{\mu m}) without the
      need for a~detailed degradation correction,  a proxy model (hereafter referred to as SCIAMACHY proxy model)
      was developed by fitting solar proxy time series to observed SCIAMACHY
      measurements over several 27-day solar rotation periods
      \citep[][]{Pagaranetal:2009}. The proxies used in this model are the
      photometric sunspot index \citep[][]{Balmacedaetal:2009} and the Mg~II
      index \citep[][]{Weberetal:2012} for sunspot darkening and facular
      brightening.  Assuming that the fitting parameters linearly scale from
      solar rotations to an 11-yr solar cycle, one can then use the solar
      proxies to extrapolate beyond the lifetime of the single instrument
      \citep[][]{Pagaranetal:2011a}. Note, however, that this assumption
      might not be accurate, and probably results in an \mbox{underestimate} of the
      magnitude of the solar cycle variation.  The short-term variability of
      the SCIAMACHY proxy model over several solar rotations agrees well
      with direct solar observations from SCIAMACHY, SORCE/SIM, and
      UARS/SUSIM \citep[][]{Pagaranetal:2009, Pagaranetal:2011b}.  Over
      longer periods during the descending phases of solar cycles 21 to 23,
      larger differences between model and direct observations become
      apparent. In particular, SORCE/SIM data show UV changes that are about
      four times larger than the SCIAMACHY proxy model 
      (\citealp{Pagaranetal:2011a}; Fig.~\ref{fig:ssicontri}).

      Very recently an optical degradation model has been developed that
      uses the various light paths from different combinations of mirrors
      within SCIAMACHY. The main causes for the optical degradation are
      believed to be contaminants on the mirror surfaces. This new
      degradation model with improved calibration corrections { should allow in the near future to derive long-term time series of SCIAMACHY measurements without the use of the SCIAMACHY proxy model.} 

\subsubsection{SORCE}
\label{sec.sorce}

      The Solar Radiation and Climate Experiment
      \citep[SORCE;][]{Rottman:2005}, launched in January 2003, has made
      continuous daily measurements of SSI from 0.1 to 2400\,\unit{nm} (with
      missing portions of the extreme UV between 35\,\unit{nm} and
      115\,\unit{nm}), accounting for about 97\,{\%} of the TSI.  These measurements are unique compared to SSI existing data. TSI is also
      measured on SORCE by the Total Irradiance Monitor
      \citep[TIM;][]{Koppetal:2005}. The two instruments onboard SORCE
      pertinent to this study are the Solar Stellar Irradiance Comparison
      Experiment \citep[SOLSTICE;][]{Mcclintocketal:2005} and the Spectral
      Irradiance Monitor \citep[SIM;][]{Harderetal:2005a, Harderetal:2005b}.

      SORCE/SOLSTICE is a~grating spectrometer that measures SSI in the UV
      from 115\,\unit{nm} to 320\,\unit{nm} with a~resolution of
      0.1\,\unit{nm} and with an absolute calibration uncertainty of
      approximately 5\,{\%} and measurement precision to better than 0.5\,{\%}
      on all time scales (Snow et~al.~2005). 
      SORCE/SOLSTICE is a~second generation of UARS/SOLSTICE
      \citep[][]{Rottmanetal:1993} which acquired UV measurements from 1991
      to 2001. SOLSTICE uses nighttime observations of stars to track and
      correct for changes in responsivity.  SOR\-CE/SOL\-STI\-CE uses two
      channels to cover the spectral regions 115--180\,\unit{nm} and
      170--320\,\unit{nm}. The long-term stability  in the
      latest data version is about 1\,{\%} per year (M.~Snow, personal communication, 2012).

      SORCE/SIM \citep[][]{Harderetal:2005a,Harderetal:2005b} was developed
      to replace the longest wavelength channel (280--420\,\unit{nm}) in the
      original UARS/SOLSTICE and to extend wavelength coverage well out into
      the NIR. SORCE/SIM employs a~single optical element, a~F\'ery prism,
      for dispersion and to focus light on four detectors in the focal
      plane. Two photodiode detectors cover the range from 200\,\unit{nm} to
      950\,\unit{nm}, another covers the range from 895\,\unit{nm} to
      1620\,\unit{nm}, and an electrical substitution radiometer (ESR)
      operates over the spectral range from 258\,\unit{nm} to
      2423\,\unit{nm}.  This ESR is also used to calibrate the other three.
      Because SORCE/SIM is a~prism spectrometer its resolution varies from
      less than 1\,\unit{nm} in UV to approximately 40\,\unit{nm} in the
      NIR.
      SORCE/SIM began reporting daily SSI results in April 2004. It  achieves  an absolute calibration uncertainty of approximately 2\,{\%}
      and measurement precision of 0.1\,{\%} or better at most
      wavelengths.  \citet[][]{Merkeletal:2011} stated that the SIM long-term uncertainty in the 200-300\,\unit{nm} region is $\approx$ 0.5-0.1\%, from 310-400\,\unit{nm} it is $\approx$ 0.2-0.05\%, and in the 400-1600\,\unit{nm} range it is better than 0.05\%. However, they also reported the lack of independent observations for direct validation of these estimates. Moreover, the values above are relative to annual changes. Therefore, they can be assumed  valid for short-term
      variations, but on longer time scales,  the instrument stability  could be considerably lower.

      It is worth mentioning that the 
      SORCE public Level~3 data  include the SOLSTICE data up to
      308\,\unit{nm} and SIM data above 308\,\unit{nm}.  However,  the SIM spectra  do extend down to about 200\,\unit{nm}.   Although the SIM data between 200-308\,\unit{nm} are not publicly available yet, they have been made exploitable 	for several studies
through personal communications.  The SORCE data used in our study to  estimate  the atmospheric response to SSI solar cycle variations  are specified in Table  \ref{table42}.   They also include data received by personal communications with instrument team members (J. Harder, personal communication, 2012).
   

      \citet[][]{Harderetal:2009} presented multi-year SORCE/SIM trends
      indicating that UV variability during the declining phase of solar
      cycle 23 (between 2004 and 2008) was larger than that observed in
      previous cycles, and was compensated by trends in other bands in the
      visible and NIR that increased with decreasing solar
      activity. SORCE/SOLSTICE has also shown enhanced UV variability for
      the same time period. Solar UV variability measured by both SORCE/SIM
      and SORCE/SOLSTICE exceed the variability observed by UARS/SOLSTICE
      and UARS/SUSIM over solar cycle 22 and ascending phase of cycle 23
      by a~factor of 3--10 depending on wavelength
      (\citealp{DelandCebula:2012}; Figs.~2,~4, and~8).

      These discrepancies with prior cycle observations and with    SSI models have
      inspired new   analyses and  collaborations aimed at a better understanding of the potential   sources of instrument
      degradation that might have affected SORCE instruments and previous
      instruments as well.   The studies  have been  concentrated on  SSI
      instrument observations, capabilities, and estimated spectral
      irradiance uncertainties, methods of correcting for degradation, and
      refinement of estimated uncertainties.  It has been understood  that all  detectors and optics suffer
      some degradation in space, largely due to exposure to solar light, and
      also due to hydrocarbon contamination that dominates below
      400\,\unit{nm}.     Accordingly, new models of degradation based on total dose, rather
      than just exposure time, are  being developed for the SORCE and other
      instruments.   Revised data sets, which e.g. will include SIM data down to 240 nm,  are expected out in 2013. Besides,
      degradation trends  have also   been analyzed by considering the expected
      invariance of SSI over the solar cycle minimum. The latter method has been
      developed by \citet[][]{Woods:2012} and applied to data during last
      solar cycle minimum (2008--2009) to estimate possible  degradation
      trends for SORCE/SIM and SORCE/SOLSTICE.   It consists of identifying
      near-identical solar activity levels on both sides of the minimum to
      derive corrections for instrument degradation. This analysis showed good  agreement of the  variability  from moderate solar activity level to minimum level from 
      various measurements and models, 
      from
120\,\unit{nm} to 300\,\unit{nm} for solar cycles 21 through 24.   However, as the method has about 30\% uncertainty in variability due  to the assumptions about selecting times of similar irradiance levels, the results may not be as accurate as those derived from  analyses based on
instrument degradation alone.

 The analysis by \citet[][]{Woods:2012} reduces the  variability of the integrated UV
      irradiance from 200\,\unit{nm} to 400\,\unit{nm}, relative to the
      measured TSI change, to  110\,{\%} (Fig. \ref{fig:ssicontri})  from the 190\,{\%}
change  reported by  \citet[][]{Harderetal:2009}.   Nevertheless, to be compatible with the observed TSI changes even this lower amplitude of the UV variation over the solar cycle still requires compensation from out-of-phase trends at other wavelengths, in particular above $\approx$ 400\,\unit{nm}.  Other analyses of solar cycle variability
      suggests that the UV variability in the 200\,\unit{nm} to
      400\,\unit{nm} range is about 60\,{\%} of the measured TSI change
      \citep[][]{Krivovaetal:2006a,
      Pagaranetal:2009, Morrilletal:2012}.

\subsubsection{SOLSPEC}
\label{sec.solspec}

      The SOLar SPECtrum instrument \citep[SOLSPEC;][]{Thuillieretal:2009}
      is composed of three double mono\-chro\-mators (170--390\,\unit{nm},
      380--850\,\unit{nm}, 800--3000\,\unit{nm}) and a~set of lamps allowing
      corrections for aging related to the harsh space environment. The
      SOLSPEC spectrometer flew several times on the Space Shuttle and its
      twin instrument, SOSP (SOlar SPectrum), was placed on the EURECA
      (EUropean Retrieval Carrier) platform for 10\,\unit{months} (April
      1993 to January 1994). These missions have provided data to build the
      ATLAS (ATmospheric Laboratory for Applications and Science) spectra,
      specifically ATLAS-1 (March 1992) and ATLAS-3 (November 1994)
      \citep[][]{Thuillieretal:2004}, which are composites using UARS/SUSIM
      and UAR/SOLSTICE data from Lyman-$\alpha$ at 121\,\unit{nm} to
      200\,\unit{nm}, and ATLAS/SSBUV, ATLAS/SUSIM, and ATLAS/SOLSPEC from
      200 to 400\,\unit{nm}, ATLAS/SOLSPEC from 400 to 850\,\unit{nm}, and
      EURECA/SOSP from 800 to 2400\,\unit{nm}. The ATLAS spectra were
      calibrated to absolute radiometric scale using the blackbody of the
      Heidelberg Observatory, and tungsten and deuterium lamps calibrated by
      NIST.

      SOLSPEC has been up-graded for operations onboard the International
      Space Station (ISS) by implementing several changes in the
      electronics, optics, and \mbox{mechanisms} and by adding redundant components
      in order to generate data with proper degradation
      correction. ISS/SOLSPEC has been calibrated to an absolute scale at
      the Physikalisch-Technische Bundesanstalt (PTB) using the BB3200pg
      blackbody radiator \citep[][]{Sperfeldetal:1998}. Over the whole
      spectral range, SOLSPEC accuracy is within 2 to 3\,{\%}. The SOLSPEC
      spectrometer has been in operation since February 2008 onboard the ISS
      along with the SOL-ACES (SOLar Auto-Calibrating EUV/UV
      Spectrophotometers) instrument measuring below 150\,\unit{nm}. When
      the ISS orientation allows, ISS/SOLSPEC records the solar spectral
      irradiance. Presently, data have been obtained during the solar
      minimum preceding solar cycle 24 and at some specific periods during
      its rising phase for direct comparisons with SORCE/SIM observations
      \citep[][]{Thuillieretal:2012a}.

\subsubsection{Statistical analysis of  SSI time series}
\label{sec.stitch}
\begin{figure}
\includegraphics[width=85mm]{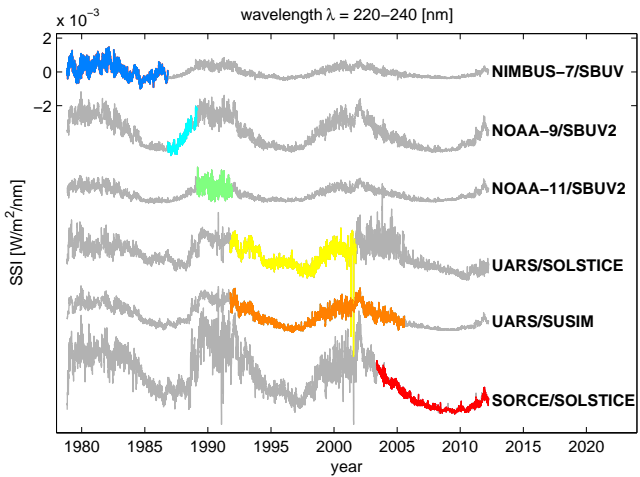}
\caption{Daily averages
of the spectral irradiance in the 220--240\,\unit{nm} band, as measured by 6
different (and only partly overlapping) instruments. Each record has
been shifted vertically for easier visualisation. Extrapolations are shown
in grey, observations in colour. One-$\sigma$ confidence intervals for
the former are not shown here but are typically 0.4\,$\times$\,10$^{-3}$ (\unit{W\,m^{-2}\,nm^{-1}}).}
\label{fig:tosca_figextrapolation}
\end{figure}

\begin{figure}
\includegraphics[width=85mm]{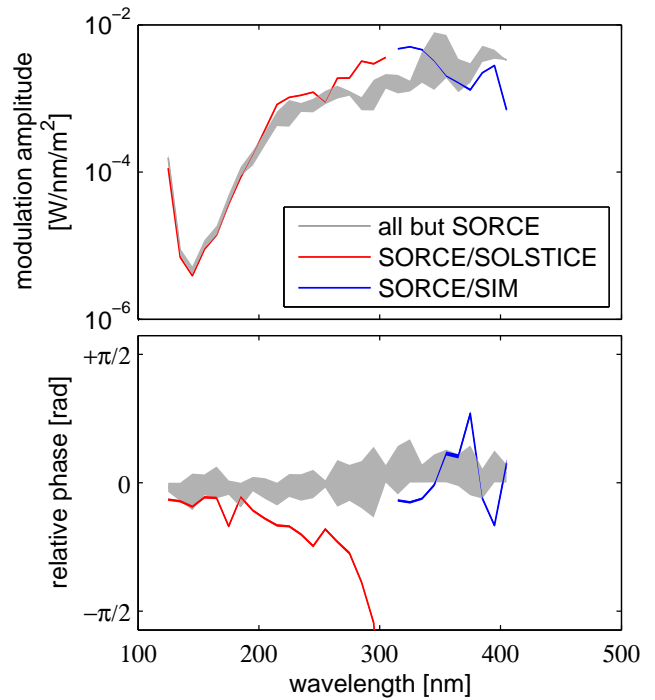}
\caption{Modulation amplitude (top) and phase (bottom) from SORCE/SIM
and SORCE/SOLSTICE, compared to that from other instruments. The SSI
has been averaged over 10\,\unit{nm} bins. The vertical width of the shaded area
 reflects the dispersion of the instruments ($\pm \sigma$ confidence interval).}
\label{fig:tosca_figmodulation}
\end{figure}

      As mentioned    above, the assessment of long-term
      variations in SSI observations can be done only by stitching together
      different records that, individually, often do not last for more than
      a~few years and are almost always offset by different calibration
      scales. The first systematic effort toward building such a~single
      composite data set for the UV was done by
      \citet[][]{DelandCebula:2008}, who created a~data set with daily
      spectra covering the wavelength range 120--400\,\unit{nm} for the time
      period November 1978 to August 2005. However, the instruments
      frequently differed in their radiometric calibration and in their
      long-term stability. One is therefore { often} left with making subjective
      adjustments that can dramatically alter the interpretation of
      long-term variations \citep[][]{Lockwood:2011} and, for example, mimic
      a~solar cycle variation that does not exist. Normalisation to a single reference spectrum, as done by \citet[][]{DelandCebula:2008}, can only partially compensate such differences in absolute calibration.  A~more objective approach
      consists of incorporating these instrumental discrepancies in the
      reconstruction, and then explicitly using them as a~contribution to
      the overall uncertainty. A~Bayesian statistical framework is { ideally suited} for
      this analysis.

      To illustrate this approach, we concentrate on the solar cycle
      (i.e. decadal) modulation, which has the advantage of being one of the
      conspicuous signatures of solar variability in climate records while
      offering sensitive means for diagnosing the quality of the solar
      observations.

Our analysis consists first in extending in time all SSI observations, and subsequently in analysing their 11-yr modulation. First, for each instrument and for each wavelength, we extrapolate the SSI observations backward and forward in time while preserving their statistical properties with respect to all other observations (cross-correlation, etc.). This is made possible thanks to the empirical evidence for all neighboring spectral bands to evolve remarkably coherently in time, on time scales of hours and beyond \citep[][]{Leanetal:1982, Amblardetal:2008}. This coherency, which is rooted in the strong magnetic coupling between solar atmospheric layers, allows us to describe all salient features of the variability in the SSI with just a few degrees of freedom. There are typically three of them in the UV. The extrapolation is based on the expectation-maximisation technique \citep[][]{Dudokdewit:2011}, which is routinely used to fill in missing data in climate records. As a result, we obtain for each wavelength as many records as there are  instruments observing it. The overall dispersion of the various observations  is then naturally reflected by the dispersion of the reconstructions, and so no offset or trend adjustments are required. We would like to highlight that no proxies are used to constrain the reconstruction.

The next step consists in comparing the    11-yr modulation amplitude and phase for each reconstruction and identifying possible changes since 1978. To date, all spectral bands and all solar proxies show in-phase variations with each other (up to a constant pass shift), regardless of the solar cycle. This is a strong, albeit not sufficient, indication that different spectral bands are likely to remain in phase from cycle to cycle. Likewise, their modulation amplitudes are unlikely to change. Any departure from this picture should thus receive special consideration. We estimate the modulation amplitudes and phases by running an 11-yr running mean through the data. Here, the phase reference is chosen to be the Mg II core-to-wing ratio because it is widely used as proxy for the solar UV \citep[][]{Vierecketal:2001}. This choice, however, has no impact on the results that follow.

      Figure~\ref{fig:tosca_figextrapolation} illustrates the extrapolation
      for the 220--240\,\unit{nm} band, which is important for ozone
      production \citep[e.g.][]{Rozanovetal:2002}. This example shows that
      all observations vary in phase with the solar cycle but differ
      considerably in their modulation amplitude. The observations from
      SORCE exhibit a~larger modulation amplitude, as already mentioned
      above.  Note that the use of the MgII index as a reference is just to fix the phase reference for comparing the results obtained for different cycles, without affecting the way the SSI is extrapolated back- and forward in time.

      Because this approach provides as many UV composites as there are
      observations, we can test whether the observations from SORCE/SIM and
      SOR\-CE/SOL\-STI\-CE are compatible with those from other instruments,
      most of which were obtained during preceding solar cycles.  The
      comparison is summarised in Fig.~\ref{fig:tosca_figmodulation},
      which compares the modulation amplitude and phase of SORCE versus the
      distribution obtained by the other instruments. We conclude that all
      instruments agree remarkably well below 200\,\unit{nm}; at longer UV
      wavelengths, the modulation amplitude inferred from 
       \mbox{SORCE/SOLSTICE}
        is
      systematically larger by a~factor of two to six at all
      wavelengths. The simultaneous sharp drop in its phase { raises doubts
      about the consistency between SOLSTICE and the other measurements, and also between SIM and the other data.}  Assuming that
      there is no reason for the SSI to be unusual during the last solar
      cycle only, we conclude that these observations are likely to be
      affected by instrumental drifts, in agreement with the conclusions
      from \citet[][]{LeanDeland:2012}. This may also be the case for
      SORCE/SIM between 308 and 340\,\unit{nm}, although the departure from
      other observations is much less significant here.

\subsection{TSI  time series}
\label{sec:tsi}

\begin{table*}[t]
\caption{Summary of  main TSI measurements from space.}
\label{table:1}
\scalebox{.9}[.9]{ 
\begin{tabular}{llllll}
\tophline
Spacecraft & Instrument & Start & End & Mean and st.dev. & Reference\\
           &            &       &     & (\unit{W\,m^{-2}})       &          \\
\middlehline
NIMBUS    & HF        & Nov~1978 & Dec~1993 & 1372.1\,$\pm$\,0.8 & \citet{Hickeyetal:1980}\\
SMM       & ACRIM-I   & Feb~1980 & Jul~1989 & 1367.5\,$\pm$\,0.7 & \citet{Willson:1979}         \\
ERBS      & ERBE      & Oct~1984 & Mar~2003 & 1365.4\,$\pm$\,0.6 & \citet{Leeetal:1987}         \\
UARS      & ACRIM-II  & Oct~1990 & Nov~2001 & 1364.4\,$\pm$\,0.5 & \citet{Papetal:1994}         \\
SOHO      & VIRGO     & Feb~1996 &          & 1365.7\,$\pm$\,0.6 & \citet{Froehlichetal:1997}   \\
ACRIM-sat & ACRIM-III & Apr~2000 &          & 1366.2\,$\pm$\,0.7 & { \citet{Willson:2001}}  \\
SORCE     & TIM       & Mar~2003 &          & 1360.9\,$\pm$\,0.4 & \citet{Lawrenceetal:2000}    \\
PICARD    & PREMOS    & Nov~2010 &          & 1360.5\,$\pm$\,0.4 & \citet{Schmutzetal:2009}     \\
\bottomhline
\end{tabular}
}
\end{table*}

   Some climate models still include  solar variability only as derived from TSI,  thereby fixing the relative spectral contribution of the solar radiation hitting the Earth.  In addition, TSI
      time series also provide constraints for the empirical and semi-empirical
      models of SSI variations    that are discussed in the following.  Therefore, it seems  appropriate to  discuss here the  current knowledge on TSI measurements.  In fact,  for  
      many
      years, the canonical value of the average TSI was
      1365.4\,$\pm$\,1.3\,\unit{W\,m^{-2}}. Now, the most accurate, and
      generally accepted, value is 1361\,$\pm$\,0.5\,\unit{W\,m^{-2}}
      \citep[][]{KoppLean:2011,Schmutzetal:2012}. This lower value will be
      used in the data assimilation and meteorological re-analysis project
      like ERA-CLIM at ECMWF (D.~Dee, ECMWF, personal communication
      2012, and see \url{http://www.era-clim.eu/}).

      TSI variability was already predicted in the 1920s from ground-based
      observations \citep[][]{Abbotetal:1923}.  Accurate measurement of TSI
      and detection of its variability requires observations from space. The
      first report of the variable solar irradiance with correct amplitudes
      was made by \citet[][]{Hickeyetal:1980}. Later observations differed
      markedly in their absolute values but all basically agreed in the
      relative amplitude of the TSI variations. In Table~\ref{table:1} we
      list the main space experiments that have measured TSI, together with
      their observed variabilities, { which for some experiments were  subsequently revised as further detailed below.} The given numbers are biased by the
      duration of the experiments and their phase relative to the solar
      cycle, but the overall result is that TSI variations are observed to
      be on the order of about 0.5\,{\%} standard deviation from the mean
      value.

      \citet[][]{Leeetal:1995} estimated the absolute accuracy of the
      NIMBUS7/HF instrument to be 0.5\,{\%} and that of ERBS/ERBE
      0.2\,{\%}. \citet[][]{Willson:1979} expected his SMM/ACRIM-I
      experiment to remain within 0.1\,{\%} for at least
      a~year. \citet[][]{FroehlichLean:1998} state that the absolute
      measurements of the early radiometers are uncertain to about
      0.4\,{\%}, which corresponds to 5.5\,\unit{W\,m^{-2}}. The main reason
      for this relatively large spread was due to the uncertainty in the
      aperture area. When the technology for determining aperture area
      improved the agreements among the measurements improved.  However, the
      SORCE/TIM experiment proved to be a~new
      outlier. \citet[][]{Lawrenceetal:2003} claim an uncertainty of
      0.5\,\unit{W\,m^{-2}}, i.e.\ accurate to 350\,\unit{ppm}. Because
      SORCE/TIM is 4.5 and 5\,\unit{W\,m^{-2}} below SOHO/VIRGO and
      ACRIM/ACRIM-III, respectively, the uncertainties given by the instrument teams do
      not overlap \citep[][]{KoppLean:2011}.

      The PREMOS experiment on the French satellite \mbox{PICARD}, which was
      launched in July 2010, has { contributed to the understanding of the instrument offsets, by confirming  the lower TSI value initially reported by the SORCE/TIM \citep[][]{Koppetal:2005}.}  The radiometers of
      the PICARD/PREMOS experiment have been calibrated in two different and
      independent ways. The first is a~calibration in power response as
      reported by \citet[][]{Schmutzetal:2009}. In addition, the TSI
      radiometer of PICARD/PREMOS is, so far, the first space instrument
      that has been calibrated in irradiance in vacuum. This was done at the
      Total solar irradiance Radiometer Facility (TRF) located at the
      Laboratory for Atmospheric and Space Physics (LASP) in Boulder,
      Colorado, USA \citep[][]{Fehlmannetal:2012}. The irradiance
      calibration is accurate to 330\,\unit{ppm}. PICARD/PREMOS agrees with
      SORCE/TIM to within 0.4\,\unit{W\,m^{-2}}, with \mbox{PICARD/PREMOS}
      being lower \citep[][]{Schmutzetal:2012}. Thus, the new experiments
      are  well within their common uncertainty range and the uncertainty
      difference between independent measurements of the { TSI}  has
      decreased by a~factor of ten. 
      
      The  new experiments and the work carried out  by the international  community, which included realization of new facilities,  ground-based tests, and collaborations aimed at identifying, quantifying, and verifying the causes of the discrepancy between the TIM and older TSI instruments, have ultimately led to the understanding of the instrument offsets.  In particular,  the characterisation of the SORCE/TIM, ACRIM/ACRIM III, and SOHO/VIRGO
      witness units at TRF, and the calibration of PICARD/PREMOS, led to better characterise instrumental effects and to resolve
      the source of the discrepancy among TSI observations
      \citep[][]{KoppLean:2011,Schmutzetal:2012}.  Instruments such as PMO6
      and ACRIM type having a~view-limiting aperture in front and a~smaller
      precision aperture that defines the irradiance area have a~large
      amount of scattered light within the instrument. This additional light
      is not fully absorbed by the baffle system and produces scattered
      light contributing to extra power measured by the cavity. Scattered
      light was one of the potential systematic errors suspected by
      \citet[][]{Butleretal:2008}. Subsequent ground testing involving the
      different instrument teams verified scattering as the primary cause of
      the discrepancy between the TIM measurements and the erroneously high
      values of other TSI instruments.  Accordingly,  new stray light corrections have recently been
      assigned to ACRIM/ACRIM-III (based on spare instruments) and its stray
      light contribution is indeed of the order as the observed
      differences. For SOHO/VIRGO the scattered light issue was not the
      reason for its discrepant reading. VIRGO is traceable to the World
      Radiometric Reference (WRR).  New analyses, however, have 
      revealed that the WRR has a~systematic offset
      \citep[][]{Fehlmannetal:2012}.  In particular, the WRR offset produced approximately the same systematic shift as the scattering error. Thus, scattered light could be 
 the reason of discrepant readings in both VIRGO and WRR instruments.

It is worth emphasizing that  the key issue in irradiance measurements is the traceability of the instrument, which is necessary to meet metrological requirements. The various instruments have undergone increasingly precise post-launch corrections to meet such requirements; PICARD/PREMOS was the first to achieve complete traceability.    However,  there is now a consistent evaluation of the TSI from four instruments: SORCE/TIM,
SOHO/VIRGO (corrected), ACRIM/ACRIM-III (corrected), and from PICARD/PREMOS.  At the same time, 
      TSI changes can be measured to much higher precision than absolute TSI
      values.  On short time scales, relative measurements are accurate to
      a~few ppm on a~daily average. On longer time scales of years and tens
      of years, the stability of the measurements are much more difficult to
      evaluate. Claims of stabilities of less than 100\,\unit{ppm} over ten
      years \citep[][]{Froehlich:2009} are most likely too
      optimistic. A~more realistic estimate comes from comparing independent
      composites that have been constructed. Over the time of the last solar
      cycle these agree to within about 0.2\,\unit{W\,m^{-2}} or about
      20\,\unit{ppm\,yr^{-1}}. For pre-1996 measurements even higher
      uncertainties for the systematic drifts have to be adopted.  Despite
      these conservative assessments, the TSI time record is a~factor of ten
      more accurate than any SSI observation.

\subsection{Discussion of SSI observations}
\label{sec.24}
      

    While the analysis of possible degradation trends for SORCE, SCIAMACHY,  and other missions  may help to resolve some of the
      differences in existing long-term time-series of measurements, to advance our understanding of SSI variability new and
      improved, upgraded observations are required in the coming years. To this purpose, the next
      generation SIM instrument  built for the NOAA/NASA Total Solar
      Irradiance Sensor (TSIS) mission \citep[][]{Cahalanetal:2012} includes many design improvements
      for reducing noise and improving in-flight degradation tracking. The
      TSIS/SIM instrument is currently undergoing laboratory calibrations  using
      radiometric  technology similar to that employed to 
      resolve the source of
      offsets among various TSI instruments \citep[][]{KoppLean:2011}. 
      
      The TSIS mission might be launched in 2016. It  is highly unlikely that TSIS/SIM and SORCE/SIM observations
      will overlap in time due to the expected lifetime of the SORCE
      batteries.  Nevertheless, even without any overlap in time, the next generation measurements will help to better understand the  performance of previous instruments and SSI solar cycle variation.  In addition, 
several satellite missions 
      dedicated to the observation of the Earth's atmosphere (GOME-2, SBUV/2, OMPS) will  also provide information on part of the SSI spectra during the possible gap in observations from the
      instruments  specifically designed to SSI measurements.

\section{Models of SSI variations}
\label{sec:models}

      Longer, uninterrupted, more stable
      and reliable observational time series are critical for understanding
      the origin of the differences between SSI measurements and improving
      our knowledge of SSI variations.  However, the physics of the
      underlying processes also needs to be understood better, in order to
      facilitate the construction of more realistic models of SSI
      variations.  Such models are particularly crucial since climate
      studies including stratospheric chemistry urgently need long and
      reliable SSI data sets for realistic simulations.  Acquiring
      sufficiently long SSI time series is a~long process and making them
      more reliable requires flying multiple new instruments, which will not
      happen for some time.  Even when such time series do become available,
      they can only be extended into the past or future with the help of
      suitable models.

      Although considerable progress has been made in modeling the TSI
      variability
      \citep[e.g.][]{FoukalLean:1990,Chapmanetal:1996,FroehlichLean:1997,
      Fliggeetal:2000a,Premingeretal:2002,Ermollietal:2003,
      Krivovaetal:2003a,Wenzleretal:2004a,Wenzleretal:2005a,
      Wenzleretal:2006a,Crouchetal:2008,Bolducetal:2012,Balletal:2012a}, modelling the SSI is more difficult,
      leaving considerable room for improvement.  Here we describe recent
      progress in SSI modelling. 
     We discuss the mechanisms responsible for the
      irradiance variations      (Sect.~\ref{sec:mecha}), with special emphasis on the possible
      differences in the spectral response of the irradiance to  the various solar magnetic features.
  We then describe the
      basic principles and key components of the models and review five
      current SSI models  (Sect.~\ref{sec:mod:mod})  that are available for climate studies and that were employed for our estimate of the atmospheric response to SSI  solar cycle variations.
      Finally, we compare the models to each
      other, confront them with the available observational data     presented in the previous section and discuss remaining
      uncertainties and open issues   (Sect.~\ref{sec:mod:sum}).

\subsection{Mechanisms of irradiance variations}
\label{sec:mecha}

      Although various mechanisms have been proposed to explain the
      variation of solar irradiance, it is now accepted that observed
      variations in TSI (i.e. over the last 3.5 solar cycles) are
      predominantly caused by magnetic features on the solar surface. We
      cannot rule out that on longer time scales other mechanisms play
      a~significant role, but this is beyond the scope of this paper.

      Empirically it has been known for a~long time that magnetic features
      on the solar surface are generally either dark (sunspots, pores) or
      bright (magnetic elements forming faculae and the network) when
      averaged over the solar disk.  Two questions arise from this
      observation: why are some flux tubes (the theoretical concept used to
      describe faculae and sunspots) bright, while others are dark? What
      happens to the energy flux blocked by sunspots (or equivalently, where
      does the excess energy emitted by faculae come from)?

      The strong magnetic field within both small and large magnetic flux
      tubes reduces the convective energy flux.  The vertical radiative
      energy flux in the convection zone is comparatively small and cannot
      compensate a~reduction in convective flux.  This leads to a~cooling of
      magnetic features.

      The magnetic features are evacuated due to the large internal magnetic
      pressure and horizontal balance of total (i.e. gas plus magnetic)
      pressure.  Hence, these evacuated magnetic structures are also heated
      by radiation flowing in from their dense and generally hot walls.
      This radiation efficiently heats features narrower than roughly
      250\,\unit{km}, making them brighter than the mainly field-free part
      of the photosphere, especially when seen near the limb where the
      bright walls are best visible \citep[][]{Spruit:1976,Kelleretal:2004}.
      For larger features, the radiation does not penetrate most of their
      volume (the horizontal photon mean free path is roughly
      50--100\,\unit{km}), so that features greater than roughly
      400\,\unit{km} in diameter remain dark (pores and sunspots);
      cf. \citet[][]{Grossmannetal:1994}.

      What happens with the energy that gets blocked by sunspots? According
      to \citet[][]{Spruit:1982b}, this energy gets redistributed throughout
      the convection zone and is re-emitted again slowly over its
      Kelvin-Helmholtz timescale, which exceeds the lifetime of sunspots by
      orders of magnitude. Similarly, the excess radiation coming from small
      flux tubes (which act as leaks in the solar surface, since these
      evacuated features \mbox{increase} the solar surface area from which
      radiation can escape) is also taken from the heat stored inside the
      entire convection zone.

      Total irradiance is simply the integral of SSI over wavelength. It is
      in the nature of integrals that quite different functions of
      wavelength, i.e. different SSI variations, can lead to the same
      variation in TSI. The differences in the relative change in irradiance
      at various wavelengths is given by three effects: (1)~The relative
      sensitivity of the Planck function to temperature increases rapidly
      with decreasing wavelength.  (2)~Radiation at the various wavelengths
      is emitted at different heights in the solar atmosphere. This
      influences SSI because the contrast of magnetic features relative to
      their non-magnetic surroundings is height- and hence
      wavelength-dependent.  (3)~At very short and at very long wavelengths
      (EUV and radio), the radiation comes from the upper transition region
      and corona, where the brightest commonly found sources are complete
      loops rather than just the loop foot-points (the flux tubes) as at
      almost all wavelengths in between.

      Point~3 refers to wavelengths we do not consider here because they
      interact mainly with the uppermost regions of the Earth's atmosphere
      (mesosphere and above) and hardly contribute to TSI at all. Point~2,
      however, is important, since in general the temperature in magnetic
      features drops more slowly with height in the solar photosphere and
      increases much more rapidly with height in the chromosphere than the
      solar average.  This means that radiation emitted at higher levels in
      the solar atmosphere, i.e. in the UV and in the cores of spectral
      lines, displays larger changes. Together, points~1 and~2 favour the UV
      to exhibit larger variations than in the visible and NIR. In addition
      to the fact that the continuum radiation comes from greater heights
      and the Planck function shows a~greater temperature sensitivity, the
      density of spectral lines per wavelength interval also increases very
      rapidly towards shorter wavelengths.

      Now, can magnetic features be dark in the visible, but bright in the
      UV?  This is in principle possible, if the magnetic feature is cool in
      the deep atmosphere, but hot in the upper photosphere.  For instance,
      pores qualitatively show such a~temperature profile, although there
      are no simultaneous high-resolution observations in the visible and
      the UV to decide if the temperature gradient is sufficiently extreme
      to produce such an \mbox{effect}.  But pores are relatively short lived and
      too few in number. Hence, any long-lasting global dimming in the
      visible at times of high activity,   such as that shown by the SORCE/SIM measurements, can only be produced by the small
      magnetic elements, specifically those in the network (these are much
      more numerous and more evenly distributed). However, such small-scale
      magnetic elements are unlikely candidates to produce a~decrease in the
      visible irradiance along with increased UV irradiance. Firstly,
      magnetic elements in the network are bright even in the continuum and
      at disk centre \citep[e.g.][]{Kobeletal:2011}.  Secondly,
      \citet[][]{Roehrbeinetal:2011} have shown that the darker than average
      appearance of some magnetic elements is largely due to spatial
      smearing of the observations (although there may be some darkening due
      to the inhibition of convection around magnetic features).  Thirdly,
      there are also many spectral lines in the visible, which brighten
      significantly in magnetic elements and counteract any darkening in the
      continuum.  Finally and most importantly, magnetic elements near the
      limb are always rather bright, so that averaged over the solar disk
      small-scale magnetic features are expected to lead to a~brightening.  All this
 implies that the measurements by SORCE/SIM of a~change in the
      irradiance at visible wavelengths out-of-phase with TSI
      \citep[][]{Harderetal:2009} are not compatible with  the surface  magnetic field
      as the source of the SSI variations in the visible.  However, the SSI
      variations in the visible are an important contributor to TSI
      variations (Fig.~\ref{fig:ssicontri}) and the above
      incompatibility would be strongly contradicted by the result of
      \citet[][]{Balletal:2012a} that 92\,{\%} of TSI variations are
      reproduced by the evolution of the magnetic field at the solar
      surface.

\subsection{Solar irradiance models}
\label{sec:mod:mod}
\begin{figure}
\includegraphics[width=85mm]{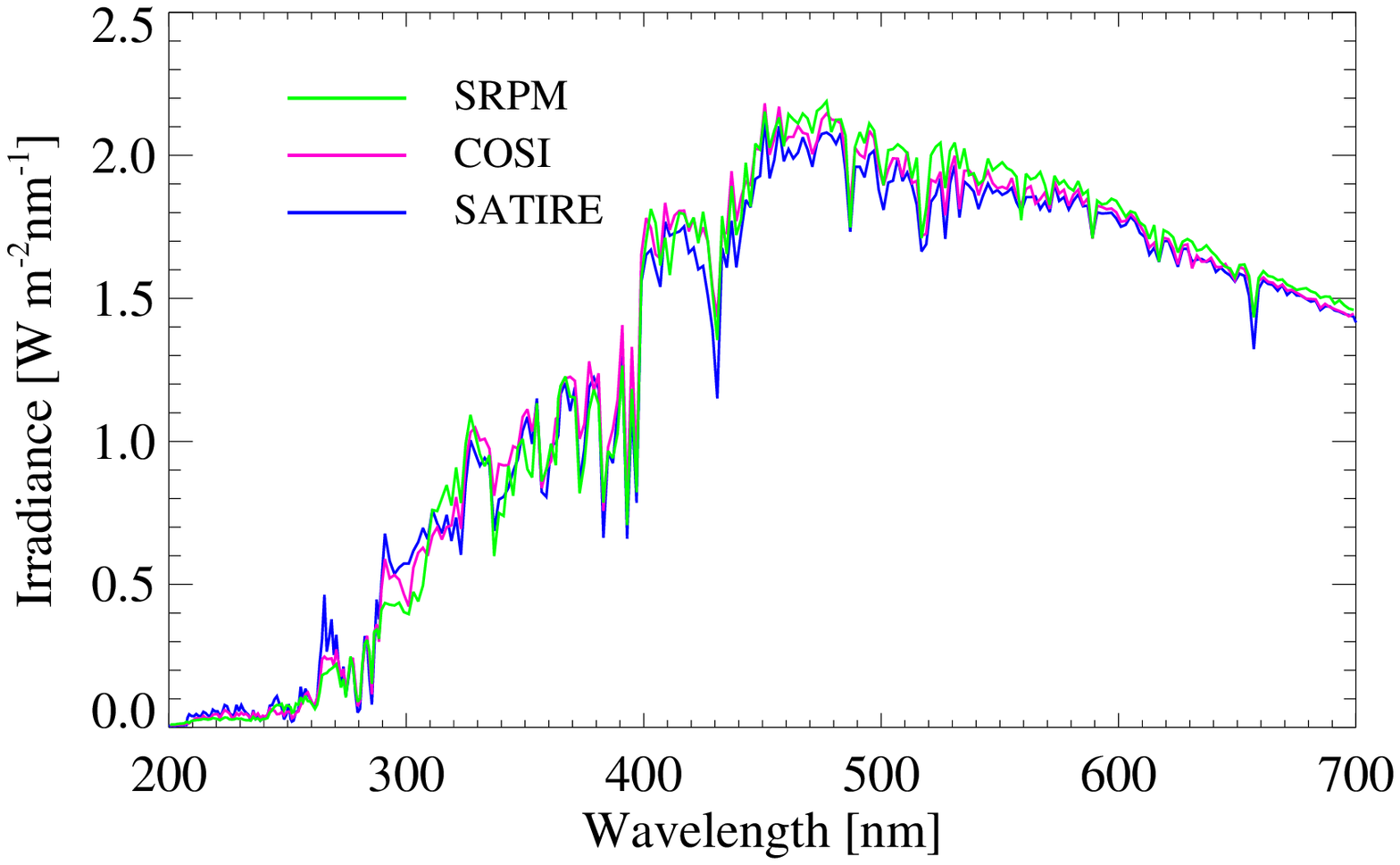}\\
\includegraphics[width=85mm]{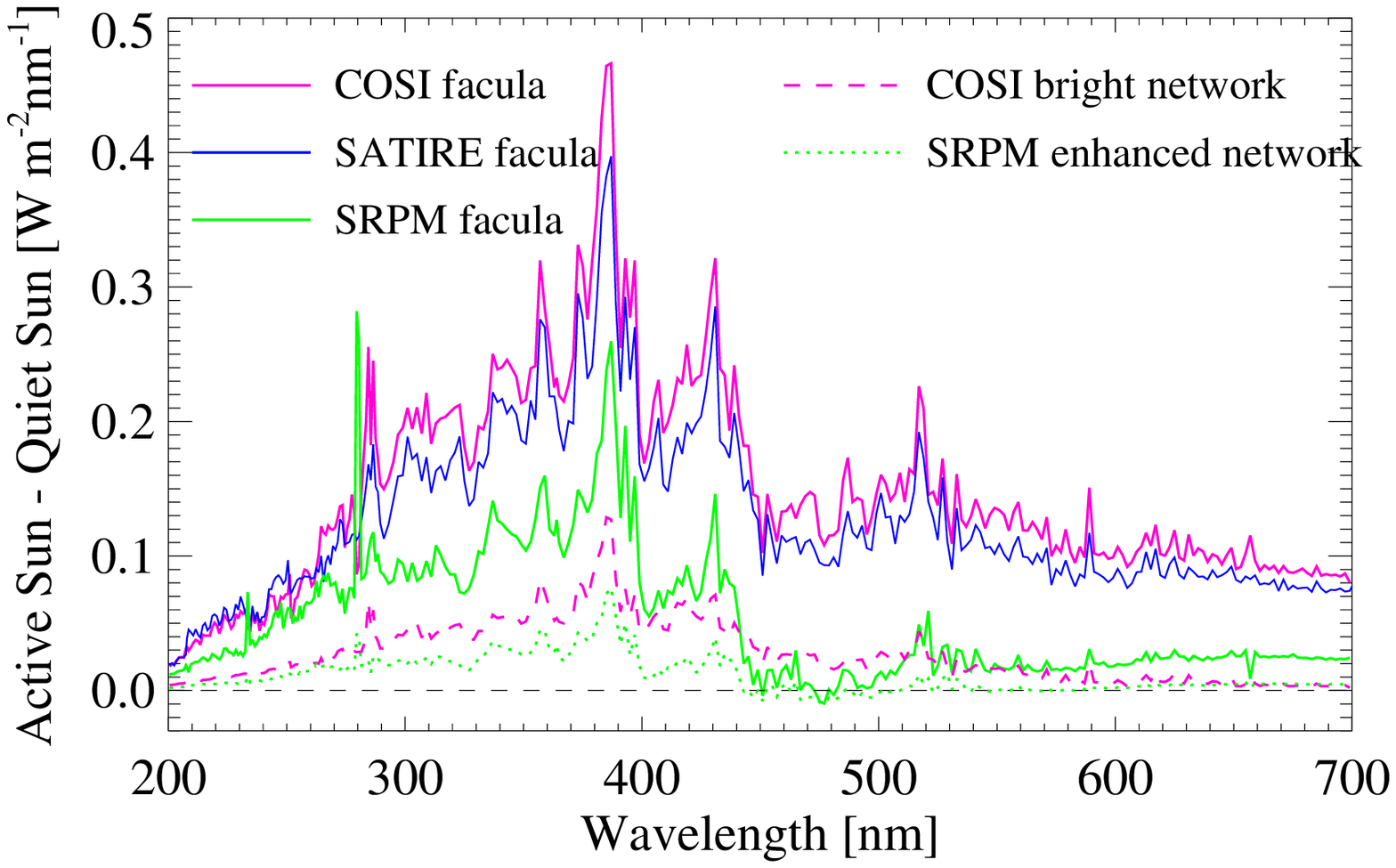}
\caption{Brightness of the quiet Sun (QS; top panel) and
brightness contrasts of
bright active components (bottom panel) in different models: SRPM \citep[][green
lines]{Fontenlaetal:2011}, COSI \citep[][purple]{Shapiroetal:2010} and SATIRE
\citep[][blue]{Unruhetal:1999}.
Note that contrasts depend on the position on the disc.
Shown are averages over the entire disc under the assumption of
a~homogeneous spatial distribution of features.}
\label{fig:contrast}
\end{figure}

      It was noticed soon after the beginning of routine monitoring of TSI
      from space, that changes in TSI were closely related to the evolution
      of different brightness structures on the visible solar disc
      \citep[][]{FoukalVernazza:1979,Willsonetal:1981,Osteretal:1982,Eddyetal:1982,FoukalLean:1986}.
      These brightness structures (such as sunspots, pores, faculae, plage
      and network) are manifestations of the solar magnetic field emerging
      at the Sun's surface.  Thus their
      evolution in a~global sense (and without looking too closely at the
      details of the temporal evolution) can be relatively well represented
      by different, typically disc-integrated, proxies of solar magnetic
      activity, such as the sunspot number or area, plage area, the solar
      radio flux at 10.7\,\unit{cm} (f10.7), the Mg~II core-to-wing index,
      or the Ca II K line.  This has widely been used in the oldest (but
      still widely deployed) irradiance models
      \cite[e.g,][]{Donnellyetal:1982,FoukalLean:1990,Chapmanetal:1996,Chapmanetal:2012,FroehlichLean:1997,Fliggeetal:1998a,Lean:2000,Premingeretal:2002,JainHasan:2004b,Pagaranetal:2009}. In
      these models, the measured irradiance variations are fitted via a~set
      of activity proxies through multiple regressions.

      The success and the limitations of the regression methods in
      accounting for measured TSI variations (on time-scales of days to
      years) gave a~strong impetus to the development of more sophisticated
      and physics-based models. Such models consider contributions of
      different brightness structures to the irradiance change separately.
      Thus the solar energy output is the sum of the fluxes emerging from
      all the features observed on the solar visible surface (corresponding
      to the solar photosphere); the number and type of disk features
      accounted for depends on the model.  Usually these models require two
      prime ingredients: (1)~the surface area covered by each photospheric
      component as a~function of time, and (2)~the brightness of each
      component as a~function of wavelength and often also of the position
      on the solar disc.

      (1)~The surface area coverage by each photospheric component (i.e.
      sunspot umbrae and penumbrae, pores, faculae, plage, network etc.) is
      derived from observations.  These could be disc-integrated data
      \citep[e.g. sunspot number, sunspot area, facular or plage area, Mg~II
      index, cosmogenic isotope data
      etc.;   such as in the models by][]{Fliggeetal:1998a,SolankiUnruh:1998,FliggeSolanki:2000,Krivovaetal:2007a,Krivovaetal:2010a,Shapiroetal:2011,Vieiraetal:2011}
      or spatially resolved observations
      \citep[ e.g. in the models by][]{Fliggeetal:2000a,Ermollietal:2003,Krivovaetal:2003a,Fontenlaetal:2004,Fontenlaetal:2011,FontenlaHarder:2005,Wenzleretal:2004a,Wenzleretal:2005a,Wenzleretal:2006a,Unruhetal:2008,Balletal:2011}.

      (2)~The brightness of individual photospheric components is calculated from semi-empirical models of the solar atmospheric structure
      using various radiative transfer codes, such as SRPM
     \citep[][and references therein]{Fontenlaetal:2011}, ATLAS9  \citep[][]{Kurucz:1993} in SATIRE  \citep[][]{Unruhetal:1999,FliggeSolanki:2000,Krivovaetal:2003a}
     or COSI \citep[][]{Haberreiteretal:2008,Shapiroetal:2010}. Such
      computations are significantly complicated by the departures from the
      local thermodynamical equilibrium (LTE) conditions \citep[Sect.  2.6
      of][]{Rutten:2003} in the solar atmosphere, as well as by the temporal
      and spatial bifurcations of the temperature and density in the solar
      atmosphere \citep[][]{CarlssonStein:1997,UitenbroekCriscuoli:2011}.
      High resolution 3-D models \citep[e.g.][]{SocasNavarro:2011} and 3-D
      MHD simulations \citep[e.g.][]{Vogler:2005} of the solar atmosphere
      are gradually becoming available.  However, current models cannot yet
      reproduce available observations over the entire spectrum
      \citep[][]{Afram:2011}. Thus, at present semi-empirical 1-D models of
      the solar atmosphere are a~de-facto standard for calculating
      irradiance variations.  Although these models do not account for the
      spatial structure and temporal variability of the solar atmosphere,
      they can be  adjusted to calculate the solar spectrum and its
      variability with high accuracy
      \citep[][and references therein]{Unruhetal:1999,Fontenlaetal:2011}.  These models do not
      necessarily catch the average properties of the inhomogeneous solar
      atmosphere \citep[][]{Loukitchevaetal:2004,UitenbroekCriscuoli:2011},
      but, if validated and constrained by the available measurements, they
      can be considered as a~reliable and convenient semi-empirical tool for
      modelling SSI variability.

      Presently, a~wide range of solar atmospheric models with different
      degrees of complexity is available
      \citep[e.g.][]{Kurucz:1993,Fontenlaetal:1999,Fontenlaetal:2009,Fontenlaetal:2011,Unruhetal:1999,Penzaetal:2004,Uitenbroek:2001,Kurucz:2005,AvrettLoeser:2008,Shapiroetal:2010},
      most of which go back to \citet[][]{Vernazzaetal:1981}.

      The spectra of the quiet Sun calculated with three different models
      are shown in Fig.~\ref{fig:contrast} (top panel).  All plotted spectra
      are in reasonable agreement with each other, though models used in
      COSI and SATIRE are a~bit closer to each other than to SRPM. The
      bottom panel of Fig.~\ref{fig:contrast} shows the flux differences
      (i.e. the contrasts) between bright active components and the quiet
      Sun calculated with different models.  The calculations were done
      assuming that the active regions cover the entire solar disk.  To
      calculate the variability, the contrasts have to be weighted by the
      filling factors (surface area \mbox{coverage}), which are specific for every
      model.  Thus, different magnitudes of the contrasts do not necessarily
      imply that the models produce different variability.  At the same
      time, the spectral profile of the contrasts determines the dependence
      of the variability on the wavelength.   Figure~\ref{fig:contrast} shows that the (disc$-$integrated) spectral contrasts of facular, plage, and network regions
      calculated with SATIRE and COSI are very similar {   and positive over the whole spectral range. On the other hand, the SRPM  contrasts are much lower 
 in the visible compared to  other models and 
 are  negative  at  some wavelengths.  This is not very
      surprising,  since the SRPM contrasts shown in
      Fig.~\ref{fig:contrast} are derived from the most recent versions of the
      \citet[][]{Fontenlaetal:1999} model family that were subsequently revised to account for some  observations of the solar atmosphere, as further detailed below.} 

      The advantage of employing radiances computed from the semi-empirical
      model atmospheres is that they allow computations of solar irradiance
      at different wavelength (i.e.  spectral irradiance).  This is not
      straightforward with the regression models, since in this case the
      regression coefficients need to be estimated from observations or
      alternatively, irradiance changes at individual wavelengths need to be
      somehow scaled from the TSI changes or changes in the irradiance at
      some other (known) wavelength.  This latter technique is also partly
      used in the UV by the NRLSSI and SATIRE models     that are   described below.

\subsubsection{NRLSSI}
\begin{figure}
\includegraphics[width=85mm]{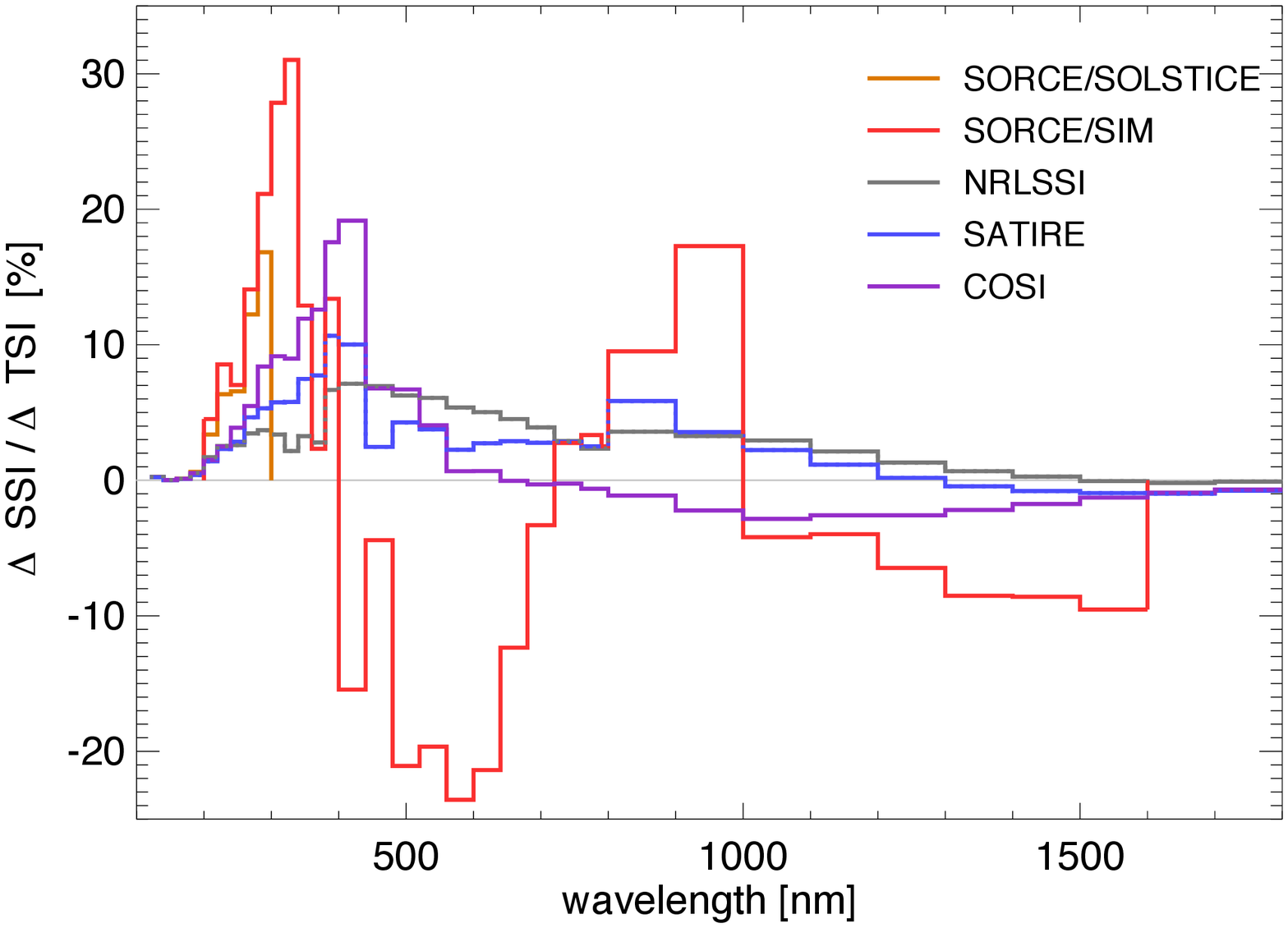}
\caption{Ratio of SSI variability to TSI variability between 200\,\unit{nm} and
1800\,\unit{nm} in bins of 20\,\unit{nm} below 400\,\unit{nm}, 40\,\unit{nm} in the range 400--800\,\unit{nm}, and
100\,\unit{nm} in the IR above 800\,\unit{nm}. Shown are SORCE/SOLSTICE (orange) and SORCE/SIM
(red) measurements between 2004 and 2008, as well as the NRLSSI (grey),
SATIRE-S (blue) and COSI (purple) models between the maximum and minimum
of cycle 23.}
\label{fig:varia}
\end{figure}

      The Naval Research Laboratory Solar Spectral
      Irradiance\footnote{http://lasp.colorado.edu/lisird/nrlssi/}
      \citep[NRLSSI;][]{Leanetal:1997,Lean:2000} uses the photospheric
      sunspot index derived from sunspot area records to describe the
      evolution of sunspots in time, and Mg II, CaII and f10.7
      disk-integrated indices to represent facular brightening.

      Below 400\,\unit{nm}, the spectral irradiances are derived from
      UARS/SOLSTICE observations through a~multiple regression analysis with
      respect to a~(SOLSTICE) reference spectrum.  The regression analysis
      includes a~facular brightening and a~sunspot \mbox{blocking} component (see
      \citealp{Leanetal:1997,Lean:2000} for more detail).  It is known that
      the long-term stability of the UARS spectral instruments (both
      SOLSTICE and SUSIM) was not sufficient to trace solar cycle
      variability above roughly 220\,\unit{nm} \citep[][]{Woodsetal:1996}.
      Therefore the coefficients are derived from the rotational variability
      so as to avoid any long-term instrumental effects.  This approach thus
      assumes that spectral irradiance changes show the same linear scaling
      with a~given proxy on rotational as well as cyclical time scales.

      Above 400\,\unit{nm}, the facular and sunspot contrasts are largely
      based on the contrasts presented in \citet[][]{SolankiUnruh:1998}.
      They have been scaled to ensure that the overall
      (wavelength-integrated) solar cycle change due to sunspots (viz
      faculae) agrees with the bolometric value for the sunspot blocking
      (viz facular brightening) derived from TSI modelling \citep[see,
      e.g.][]{FroehlichLean:2004}.

      The quiet-Sun spectrum in NRLSSI is effectively a~composite of
      UARS/SOLSTICE observations (below 400\,\unit{nm}), SOLSPEC
      \citep[][]{Thuillieretal:1998} and the model by
      \citet[][]{Kurucz:1991}. This composite has been scaled so that its
      integrated flux corresponds to a~TSI value of 1365.5\,\unit{W\,m^{-2}}.

       Compared to the SATIRE and COSI models (described below), the NRLSSI model shows lower variations on solar-cycle and longer time scales between 250 and 400\,\unit{nm}  (Figs.~\ref{fig:ssicontri} and~\ref{fig:varia}). This is mainly because the regression coefficients are derived from rotational variability only.
      Below 250\,\unit{nm} (Figs.~\ref{fig:varia} and~\ref{fig:refrange})
      and in the visible at 400--700\,\unit{nm} (Figs.~\ref{fig:ssicontri}
      and~\ref{fig:varia}) the NRLSSI, COSI, and SATIRE  models agree well with each other.

      In the visible and IR range, the facular contrasts in the NRLSSI model
      rely on the facular model atmosphere by \citet[][]{SolankiUnruh:1998},
      which is an earlier version of the model by \citet[][]{Unruhetal:1999}
      currently used in SATIRE.  Thus although the NRLSSI model does show
      a~weak out-of-phase variability in the IR around 1600\,\unit{nm}
      (Fig.~\ref{fig:varia}), integrated over the range
      1000--2430\,\unit{nm} (Fig.~\ref{fig:ssicontri}) the modelled
      irradiance is in phase with the solar cycle.  The SATIRE, COSI and OAR
      models all display reversed variability in this range.  Such
      a~reversed behaviour is also suggested by the SORCE/SIM data, although
      the magnitude of the measured SIM variability is much stronger than in
      the models.

      Integrated over all wavelengths, the NRLSSI irradiance (i.e. the TSI)
      is in good agreement with the measurements, although it does not quite
      reproduce the comparatively low TSI level during the last minimum in
      2009 \citep[][]{KoppLean:2011}, as indicated by the PMOD composite and
      also independently found by \citet[][]{Balletal:2012a} with the SATIRE
      model.  This is because the Mg~II index employed to describe the
      evolution of the bright component with time does not follow exactly
      the shape of the TSI variation over the last cycle (although the
      differences appear to be within the long-term measurement
      uncertainties; M.~Snow, personal communication, 2012).

\subsubsection{SATIRE-S}

\begin{figure}
\includegraphics[width=85mm]{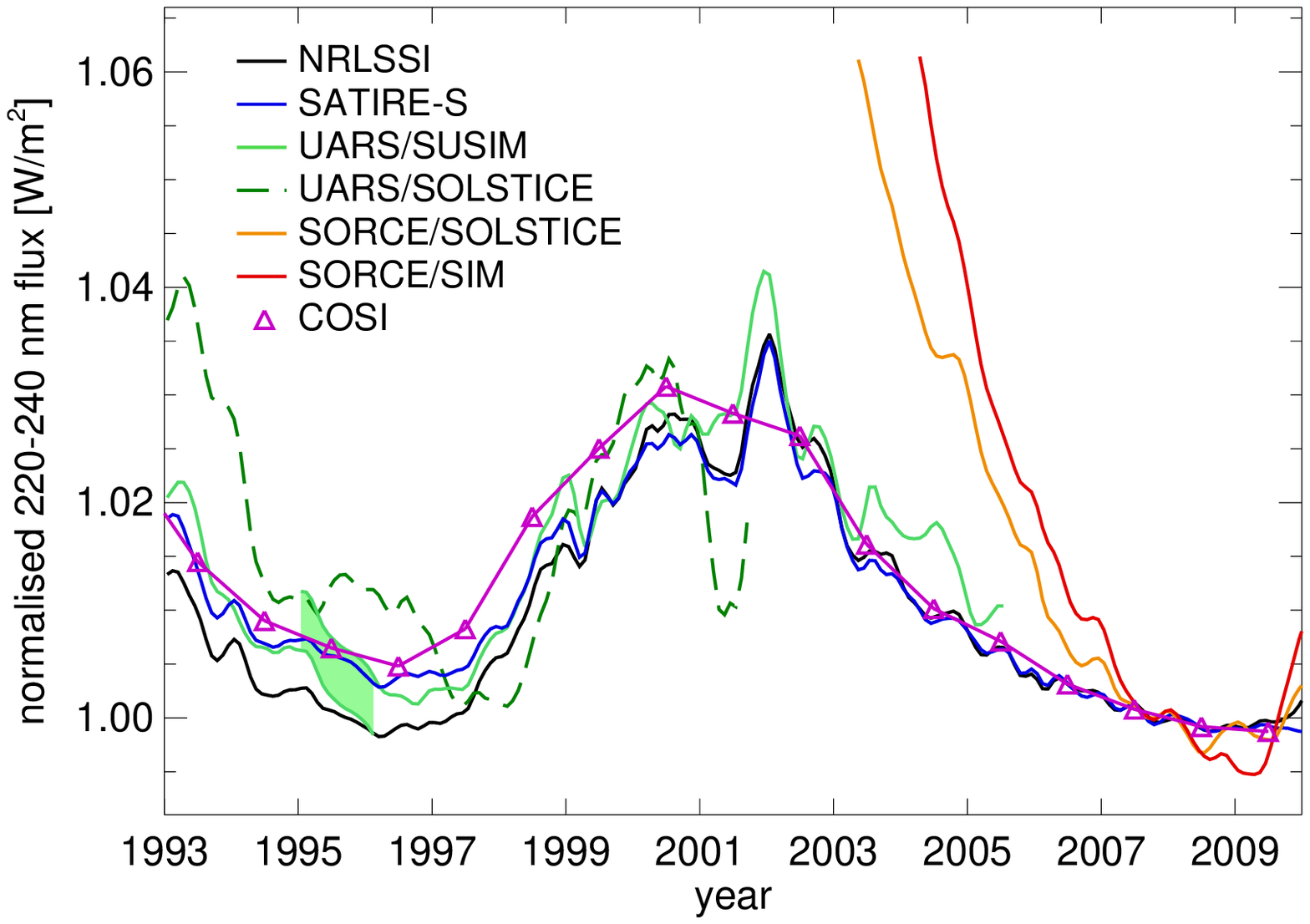}\\
\includegraphics[width=85mm]{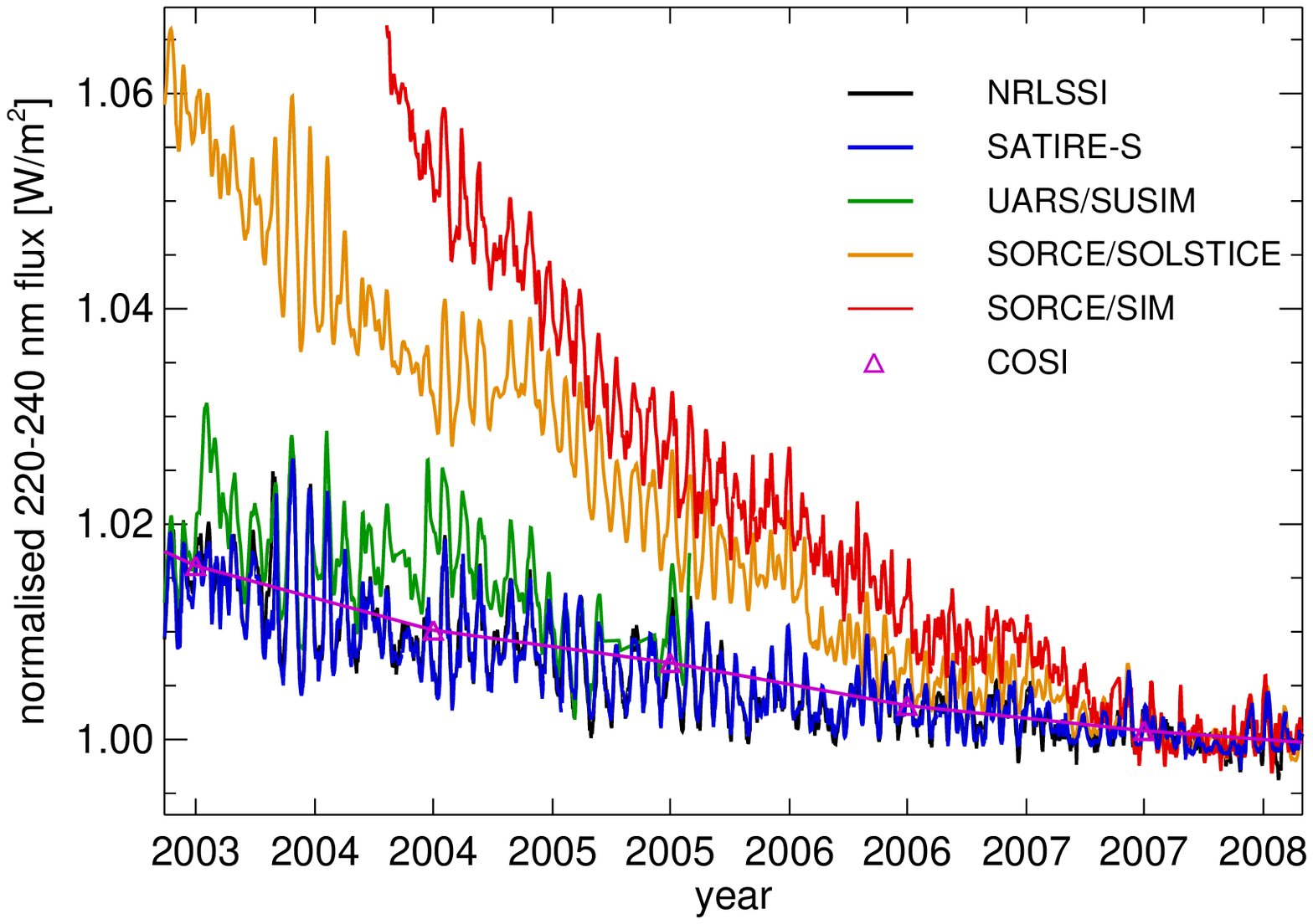}
\caption{Solar UV irradiance between 220 and 240\,\unit{nm} calculated
with NRLSSI (black), SATIRE-S (blue) and COSI (magenta), and measured
with UARS/SUSIM (darker green), UARS/SOLSTICE (light green), SORCE/SOLSTICE
(orange) and SORCE/SIM (red).
The pale green shading in the upper panel
marks the period when the sensitivity of the
UARS/SUSIM instrument (and thus the flux) changed, so that a~shift was
applied to the data before that (see
\citealp{Krivovaetal:2006a,Krivovaetal:2009a} for details).
Top panel shows 3-month smoothed values over the period 1993--2009, while the
bottom panel is limited to the period when SORCE was in operation, i.e.
after 2003, and shows daily values, except for the COSI model, for which
only yearly averages are available. All quantities are normalised to the corresponding mean values during the last activity minimum (2009).}
\label{fig:refrange}
\end{figure}

      SATIRE-S\footnote{\url{http://www.mps.mpg.de/projects/sun-climate/data.html}}
      (Spectral And Total Irradiance REconstructions for the Satellite era)
      belongs to the class of models using semi-empirical model atmospheres
      to calculate brightnesses of different surface features and full-disc
      solar images to describe the surface area coverage by these components
      at a~given time.  The full-disc data used in SATIRE are magnetograms
      and continuum images.  So far, ground-based KP/NSO (Kitt Peak National
      Solar Observatory; 1974--2003) and space-based SOHO/MDI (Michelson
      Doppler Interferometer; 1996--2010) and SDO/HMI (Helioseismic and
      Magnetic Imager; since 2010) data have been employed.  The following
      model atmospheres are used: the quiet Sun model by
      \citet[][]{Kurucz:1991} with an effective temperature of
      5777\,\unit{K}, similar but cooler model atmospheres for umbra and
      penumbra, and model P of \citet[][]{Fontenlaetal:1999} slightly
      modified by \citet[][]{Unruhetal:1999} to achieve better agreement
      with observations in the visible and near-UV.  Since the ATLAS9 code
      uses the LTE approximation, which is known to fail in the UV, fluxes
      below 270\,\unit{nm} are re-scaled using UARS/SUSIM observations
      \citep[][]{Krivovaetal:2006a}.

      The SATIRE-S modelled variability was found to be in very good
      agreement, on both rotational and cyclic time scales, with the PMOD
      TSI composite \citep[][]{Wenzleretal:2009a,Balletal:2012a}, SORCE/TIM
      TSI measurements \citep[][]{Balletal:2011,Balletal:2012a}, UARS/SUSIM
      spectral irradiance
      \citep[][]{Krivovaetal:2006a,Krivovaetal:2009a,Unruhetal:2012}, as
      well as with UARS/SUSIM and UARS/SOLSTICE Ly-$\alpha$ measurements
      (Figs.~\ref{fig:ssicontri} and~\ref{fig:refrange}).  On
      rotational times scales, good agreement is also found between the SSI
      provided by SATIRE-S and the SORCE/SIM and SORCE/SOLSTICE measurements
      \citep[][see also bottom panel of
      Fig.~\ref{fig:refrange}]{Unruhetal:2008,Unruhetal:2012,Balletal:2011}
      and between SATIRE-S and SOHO/VIRGO observations in three spectral
      channels in the near-UV, visible and near-IR
      \citep[][]{Krivovaetal:2003a}.  Due to strong sensitivity trends,
      VIRGO spectral data could not be used on longer time scales.  On time
      scales longer than a~few months, the SATIRE-S trends, however, diverge
      significantly from those shown by SORCE/SIM and SORCE/SOLSTICE: the
      change in the UV is significantly weaker in SATIRE-S, and the inverse
      solar cycle change in the visible is not reproduced \citep[
      Figs.~\ref{fig:varia} and
      \ref{fig:refrange};][]{Balletal:2011,Unruhetal:2012}.  Interestingly,
      \citet[][]{Balletal:2011} have shown that when integrated over the
      spectral range of SIM (200--1630\,\unit{nm}) and corrected for the
      missing UV and IR wavelengths, SATIRE-S still reproduces over 94\,{\%}
      of TSI fluctuations measured by SORCE/TIM.   { The trends of smoothed data agree to over 99\,{\%}.   At the same time, wavelength$-$integrated SORCE/SIM data show different trends (only  about 60\% of the TIM changes are  reproduced over the SORCE/SIM life time), though the uncertainties in the integrated SORCE/SIM data  are quite large \citep[see][]{Balletal:2011}.}

\subsubsection{SRPM}

      The Solar Radiation Physical
      Modelling\footnote{\url{http://www.digidyna.com/Results2010/}} (SRPM) is
      a~set of tools used to construct semi-empirical models of the solar
      atmosphere   and to derive solar radiance (or emitted intensity)  and  high-resolution irradiance spectra of the Sun.    SRPM
      covers all levels of the solar atmosphere from the photosphere to the
      corona and    takes, for most species, NLTE conditions into account. The calculations include a large number of atomic levels and lines, as well as molecular lines and molecular photo-dissociation opacities.      Seven components have been used in SRPM until recently corresponding
      to different features that can be identified on the Sun at
      a~medium-resolution of $\approx$\,2\,\unit{arc\,sec}: 
      quiet-Sun
      inter-network, quiet-Sun network lane, enhanced network, plage (that
      is not facula), facula (i.e. very bright plage), sunspot umbra and
      sunspot penumbra.      The atmospheric models adopted for the calculations of each of these features are presented in   \citet[][]{Fontenlaetal:2009}.  These  models go back to the ones of \citet[][and references therein]{Fontenlaetal:1999}, which have been revised in subsequent studies of \citet[][]{Fontenlaetal:2004,Fontenlaetal:2006,Fontenlaetal:2009} to account for observational results and  for extension of the atmospheric layers. The observational results include e.g. the negative contrast measurements on facular regions at some wavelengths  \citep[][]{Foukaletal:1990,Topkaetal:1992,Wangetal:1998}). 
Recently, further modifications were introduced into the plage models to take SORCE data into account (see below). In addition, two new 
components and      corresponding model atmospheres, namely dark quiet-Sun inter-network and hot facula, were added
but not yet integrated into solar irradiance calculations presented in
\citet[][]{Fontenlaetal:2011}. 
   For irradiance calculations,  the
      distribution of the atmospheric components over the solar disc at
      a~given time is derived from full-disc solar filtergrams; the ones employed so far include observations
      from the Precision Solar Photometric Telescopes
      \citep[PSPT;][]{CoulterKuhn:1994} at the INAF Osservatorio Astronomico
      di Roma 
      and at Mauna Loa, as well as Meudon photographic images.  
      
     \citet[][]{Fontenlaetal:2011}  presented   SRPM-based calculations of daily SSI at 1\,\unit{nm} resolution 
 for the  period 2000-2009 using  observations  carried out with the PSPT telescope in Rome. 
 The atmospheric model set used in \citet[][]{Fontenlaetal:2011} is essentially that of \citet[][]{Fontenlaetal:2009} with a few modifications, e.g.  to photospheric layers of the plage models  \citep[see][]{Fontenlaetal:2011}.These 
 calculations produce a decrease in irradiance  at wavelengths below $\approx$400\,\unit{nm} during the declining phase of solar activity from 2002 to 2009,  and a flux increase at longer wavelengths, similarly to the behaviour seen in the SORCE/SIM observations \ \citep[][]{Harderetal:2009}.  The UV variability derived from the model is,   however,  significantly weaker than what is  measured by SORCE/SOLSTICE. Moreover, the 
observed changes of TSI on time scales longer than the
solar rotation are not captured by the model.  
  \citet[][]{Fontenlaetal:2011}  argue that a possible reason for this mismatch could be that the current set of the  employed atmospheric models is still missing some components.

\subsubsection{COSI}

      The COde for Solar
      Irradiance\footnote{\url{ftp://ftp.pmodwrc.ch/pub/Sasha/}} (COSI)
      calculates synthetic solar spectra for different components of the
      solar atmosphere.  COSI returns the most important UV lines including
      the degree of ionisation of the elements under NLTE.  
      { The NLTE Opacity
      \mbox{Distribution} Functions \citep[][]{Hubenyetal:1995}, implemented in COSI by \citet[][]{Haberreiteretal:2008}}, indirectly account for
      the NLTE effects in several millions of lines.
     \cite{Shapiroetal:2010} showed that NLTE effects influence the
      concentration of the negative ion of hydrogen which results in an
      approximately 10\,{\%} change of the continuum level in the visible
      spectrum.  The contrasts between active regions and the quiet Sun are
      also affected by the NLTE effects.  This, however, does not imply that
      only NLTE codes are capable of reproducing the visible solar spectrum
      as the NLTE effects can be imitated by a~slight readjustment of the
      atmosphere structure \citep[see,
      e.g.][]{RuttenKostik:1982,ShchukinaTrujillo:2001}.  The spectrum of
      the quiet Sun calculated with COSI is in good agreement with the solar
      spectrum measured by SOLSTICE (up to 320\,\unit{nm}) and SIM (above
      320\,\unit{nm}) onboard the SORCE satellite during the 2008 solar
      minimum, and with the SOLSPEC measurements during the ATLAS~3 mission
      in 1994 \citep[][]{Thuillieretal:2011}.  \citet[][]{Shapiroetal:2012a}
      showed that COSI can accurately reproduce the center-to-limb variation
      of solar brightness in the Herzberg continuum retrieved from the
      analysis of solar eclipses observed by LYRA/PROBA2.

      Solar spectral irradiance variations can be modelled from COSI spectra
      by weighting them with filling factors of surface magnetic features
      (e.g.  derived from the magnetograms as in the SATIRE model,  cf. \citet[][]{Haberreiteretal:2005}, or from
      the PSPT images as in SRPM).  In the current version of the model, this has been done employing
      sunspot and $^{10}$Be data \citep[][]{Shapiroetal:2011} and and PSPT images \citep[][]{Shapiroetal:2012b}.

      The variability so derived with COSI agrees well with SATIRE-S and
      NRLSSI in the Herzberg continuum spectral range (Figs.~\ref{fig:varia} and~\ref{fig:refrange}) and in the visible (Figs.~\ref{fig:ssicontri} and~\ref{fig:varia}).  In the near-IR, at
      700--1000\,\unit{nm}, COSI shows a~weak inverse solar cycle
      variability (Figs.~\ref{fig:ssicontri} and~\ref{fig:varia}).  This
      is probably an artifact of the model, as \citet[][]{Shapiroetal:2010}
      calculated active-region contrasts using the model atmospheres of
      \citet[][]{Fontenlaetal:1999}, which  do not distinguish between umbra
      and penumbra. Besides, they employed a single model for both bright plage and plage. These assumptions affect  the
      position of the inversion point; the point where the influence of
      sunspots starts to outweigh the bright components.  To 
      account for this, \citet[][]{Shapiroetal:2012b} decreased plage and sunspot
      contrasts with respect to the quiet Sun.  A~more accurate approach is
      under development.

\subsubsection{OAR}

      Full-disc observations carried out with the PSPT telescope in Rome
      \citep[][]{Ermollietal:1998,Ermollietal:2007,Ermollietal:2010} are
      also employed in the OAR (Osservatorio Astronomico di Roma) model of
      solar irradiance variations.  Earlier results obtained with both
      a~regression method and a~semi-empirical model were presented by, e.g.
      \citet[][]{Penzaetal:2003,Penzaetal:2006},
      \citet[][]{Domingoetal:2009} and \citet[][]{Ermollietal:2011}.  A~new
      semi-empirical model, briefly outlined below, is currently under
      development (Ermolli et~al., 2012).

      Images taken between September 1997 and January 2012 have been
      processed and segmented to identify different features on the solar
      disc as described by \citet[][]{Ermollietal:2010}.  The  seven classes of
      atmospheric features proposed by \citet[][]{Fontenlaetal:2009} are considered. The brightness
      spectrum of each feature is computed using the semi-empirical
      atmospheric models of \citet[][]{Fontenlaetal:2009} through the RH
      synthesis code \citep[][]{Uitenbroek:2002}.
      Thus the set of components in the OAR model is essentially the same as
      used by \citet[][]{Fontenlaetal:2011}, though their atmospheric
      structure is based on \citet[][]{Fontenlaetal:2009}, i.e.  before     the most  recent modifications applied  by \citet[][]{Fontenlaetal:2011}.
      
   The SSI
      variation calculated with the OAR model in the visible (400--691\,\unit{nm}) over the period 2000--2012 shows  a trend   that is  opposite to that measured by SORCE/SIM and obtained with  SRPM calculations. The variability in the
      UV (200--400\,\unit{nm}) estimated with the model is significantly weaker than that depicted by the SORCE measurements.   On the other hand,   the TSI
      variations calculated with the model agree well with the PMOD
      composite and SORCE/TIM measurements on both rotational and cyclic
      time scales (Ermolli et~al., 2012),    thus suggesting that there is no need for  additional components to reproduce the observed TSI changes.

\subsection{Discussion of  SSI model results}
\label{sec:mod:sum}

There has been steady progress in modelling SSI 
variations, and a~number of models are now available that can be used as input
for climate studies.
However, the main uncertainty in the models concerns the wavelength range
220--400\,\unit{nm}, which is of particular interest for climate studies. 
   
      Of the five models discussed in this section, specifically NRLSSI,
      SATIRE-S, COSI, SRPM and OAR, only one (SRPM) shows a~behaviour of the UV and visible irradiance 
      qualitatively resembling that of the recent SORCE/SIM measurements.   However, it should be noted that the integral of the  SSI computed with this model over the entire
      spectral range (i.e.  the TSI) does not reproduce the measured cyclical TSI changes.
      None of the other four models, which  are in closer agreement with each other,  reproduces the  behaviour of the UV and visible
      irradiance observed by SORCE/SIM
      (Figs.~\ref{fig:ssicontri},~\ref{fig:varia} and~\ref{fig:refrange}).
     These models are nevertheless all in good or fair agreement with
      earlier UARS UV observations, TSI measurements and the model based on
      SCIAMACHY data
      \citep[e.g.][]{Krivovaetal:2006a,Krivovaetal:2009a,Pagaranetal:2009,Morrilletal:2011,Shapiroetal:2011,Unruhetal:2012,LeanDeland:2012,Balletal:2012a}.
   These models also agree with SORCE data on rotational
      times scales \citep[e.g.][see also bottom panel of
      Fig.~\ref{fig:refrange}]{Unruhetal:2008,Balletal:2011,LeanDeland:2012}.
      Note that good agreement on rotational time scales was also found by
      \citet[][]{DelandCebula:2012} between SORCE and other spectral
      observations.

      When comparing various models and data to each other, however, one
      important issue often remains unacknowledged, namely the true
      uncertainties in the measurements, in particular on time scales of
      years and longer.  
      As an example, 
      we compare measured and modelled variability at 220--240\,\unit{nm}
      (Fig.~\ref{fig:refrange}).  Over the period 2004--2008, the three
      models shown in this figure (NRLSSI, SATIRE-S and COSI) suggest
      a~decrease in the integrated 220--240\,\unit{nm} flux of about
      1\,{\%}.  The fluxes measured by the SORCE/SOLSTICE and SORCE/SIM
      instruments in this range decreased over the same period by roughly
      4\,{\%} and 7\,{\%}, respectively. In the 220--240\,\unit{nm} region,
      the accuracy  of SORCE/SOLSTICE is considered
      to be better than that of SIM \citep[][]{Harderetal:2010}.  Currently,
      SOLSTICE long-term stability in this spectral range is estimated to be about
      1\,\unit{{\%}\,yr^{-1}} \citep[][]{Snowetal:2005}.  Thus the observed
      trend for SORCE/SOLSTICE (4\,{\%} over 5\,\unit{yr}) and its
      difference with the models (3\,{\%} over 5\,\unit{yr}), in fact, lies
      within the long-term instrumental uncertainty.  
       The difference between
      the trends for SORCE/SIM and the models (about 6\,{\%}) is just
      outside the value of 5\,{\%} over 5\,\unit{yr}, by assuming that the SIM long-term stability is comparable to that of SOLSTICE.  Although the SIM uncertainty stated in \citet[][]{Merkeletal:2011} is lower, this assumption seems to be a reasonable working hypothesis for our study, because it is known that at
      220--240\,\unit{SIM} has poorer precision than SORCE/SOLSTICE (J. Harder, personal communication, 2012). It is worth mentioning that new assessments of the SORCE instrument degradation might imply a revision of results derived from several studies, including this one, which however is mainly intended to discuss the data and models available to date for the evaluation of the atmospheric response to SSI variations.

      Despite showing much lower variability than SORCE/SIM in the UV, the
      models also display considerable differences between each other in the
      range 250--400\,\unit{nm} (up to a~factor of three, e.g. between
      NRLSSI and COSI; Figs.~\ref{fig:ssicontri} and~\ref{fig:varia}).  { As a consequence of
      the low response of UARS/SOLSTICE to long-term variability  above  300\,\unit{nm} and the use of rotational variability to estimate the regression coefficients, it is likely that NRLSSI}
      underestimates the changes in this range (J. Lean, personal communication, 2012), and can thus be considered as the lower limit.  All other
      models rely on semi-empirical model atmospheres in this range, which
      also need further tests at all wavelengths.  The differences between
      the models in the IR are mainly due to the lack of reliable
      observations of contrasts of different (solar) atmospheric components
      and related uncertainties in the corresponding model atmospheres.  In
      the near-IR (700--1000\,\unit{nm}), all models qualitatively agree,
      except COSI.  However, as  discussed earlier, the inverse variability shown by
      COSI in this range is  believed to be an artifact  of the atmospheric components and models adopted for the irradiance calculations.

      The variability of the spectral irradiance on time scales longer than
      the solar cycle is beyond the scope of this paper.  We note, however,
      that uncertainties in the long-term SSI reconstructions are
      essentially the same as the ones discussed in this section convolved
      with the uncertainty in the magnitude of the secular change in
      irradiance \citep[see, e.g.][and references
      therein]{KrivovaSolanki:2012,Schmidtetal:2012,Solankietal:2013}.

\section{Climate impact of SSI  measurements}
\label{sec:climate}

       The  SSI data derived from  recent solar observations and model calculations presented in the previous sections are employed here to discuss the impact of SSI variations on the Earth's atmosphere and climate.    In particular, we 
      present  the motivation for  simulations  of the response of the Earth's atmosphere  to variations induced by SSI changes (Sect.~\ref{sec:41}). We then 
 discuss the  solar induced differences in atmospheric heating
      rates, ozone variability, temperature, and atmospheric circulation (Sect.~\ref{sec:42}).  These differences are       investigated in recent simulations using atmospheric models with
      standard solar forcing from the NRLSSI model \citep[e.g.][for CMIP5
      simulations]{SPARC-CCMVal:2010,Tayloretal:2012}. Additionally
      a~comparison to recent atmospheric model simulations using the
       SORCE measurements as solar forcing \citep[e.g.][]{Haighetal:2010,
      Merkeletal:2011, Oberlaenderetal:2012} is provided. As shown and
      discussed in the previous sections, the NRLSSI model output and the
    SORCE measurements represent the lower and the upper boundaries of
      current  understanding of SSI solar-cycle variability.    Here we       provide an estimate  of the uncertainty in  the magnitude of solar induced atmospheric
      changes in CCMs by comparing the impact from the two  extremes
      in the current evaluation  of SSI variability  (Sect.~\ref{sec:43}). 

\subsection{Climate modelling}
\label{sec:41}

      Today, the most advanced tools available for climate simulations are
      3-dimensional General Circulation Models (GCMs) that numerically
      simulate the general circulation of the atmosphere and/or the ocean
      based on well-established physical principles. Most climate models
      that are utilised for future climate predictions in the 4th IPCC
      report \citep[][]{Solomonetal:2007} are coupled atmosphere-ocean
      models reaching up to the middle stratosphere (32\,\unit{km}), whereas
      CCMs that are used for future predictions of the stratospheric ozone
      layer in the WMO report \citep[][]{WMO:2011} include interactive
      stratospheric chemistry and reach up to the lower mesosphere or above
      (80\,\unit{km}). The main external driving force for all climate
      models is the incoming solar flux at the top of the atmosphere (TOA).
      Currently there are two mechanisms for solar radiation influence on
      climate, the so-called \qut{top-down} UV effect
      \citep[][]{KoderaKuroda:2002} and the \qut{bottom-up} TSI effect
      \citep[][]{vanLoonetal:2007}.

      As discussed in the previous sections, even  though TSI varies only by about 0.1\,{\%} over the solar cycle,
      larger variations of several percent occur in the UV part of the
      spectrum, including  the ozone absorption bands between 200 and
      400\,\unit{nm} that are responsible for the SW heating of the
      stratosphere and are important for photochemical processes
      \citep[e.g.][]{Haigh:1994}.  The   cyclic variation of the incoming solar irradiance at short
      wavelengths leads to
      statistically significant ozone, temperature and zonal wind solar
      signals in the stratosphere \citep[][]{Austinetal:2008,Grayetal:2010}.
      These solar induced circulation changes in the stratosphere can induce
      noticeable decadal climate changes in the lower atmosphere and at the
      surface \citep[e.g.][]{Haigh:1999,Kodera:2002,
      Matthesetal:2006,Inesonetal:2011,Matthes:2011}. In order to account
      for this so-called \qut{top-down} stratospheric UV mechanism, the
      radiation code in climate models has to account for spectrally
      resolved irradiance changes. The first climate models focused on
      tropospheric climate, thus solar solar forcing was represented by TSI
      only. As stratospheric changes played a~minor role for climate
      predictions in the past, climate models did not take into account
      stratospheric processes and in particular ozone changes due to solar
      UV absorption. Hence, most of the SW radiation codes developed for use
      in GCMs did not consider solar irradiance for wavelengths shorter than
      250\,\unit{nm} and employed parameterisation using TSI as
      input. Solar fluxes and heating rates were subsequently calculated in
      one or two SW absorption bands from the top of the atmosphere to the
      surface. In contrast, middle atmosphere models use SW radiation codes
      specifically designed for simulations of the upper atmosphere. They,
      therefore, cover a~broader spectral range and include more than two
      spectral bands in the UV and visible.

      Studies on the performance of SW radiation codes with different
      spectral resolution showed that the observed solar temperature signal
      in the stratosphere can only be reproduced in models that allow for
      the effects of spectral variations between solar minimum and maximum
      \citep[][]{Egorovaetal:2004,Nissenetal:2007}.  In a~recent paper,
      \citet[][]{Forsteretal:2011} examined in detail the sensitivity of
      a~number of CCM SW radiation codes to changes in solar irradiance and
      ozone as well as the ability of the models to reproduce the 11-yr
      radiative solar signal using the NRLSSI data.  In their study
      \citep[][]{SPARC-CCMVal:2010}, the strongest solar temperature signal
      was found to be in the tropical upper stratosphere/lower mesosphere,
      indicating that the direct mechanism of heating by absorption of
      enhanced UV radiation at solar maximum is well captured by the models
      that employ spectrally resolved SW radiation schemes.  Models that do
      not account for SSI variations and only consider changes in spectrally
      integrated TSI cannot properly simulate solar induced variations in
      stratospheric temperature \citep[][]{Forsteretal:2011}.

      Today stratospheric processes are gaining a~lot of interest due to
      their importance for climate. Not only the effect of ozone recovery
      and its relationship to climate but also stratosphere-troposphere
      dynamical coupling and its role for predictability from days to
      decades have been recognised as important issues for future climate
      studies \citep[e.g.][]{Baldwinetal:2007,Gerberetal:2010}. Therefore,
      a~better representation of the stratosphere including improved
      representation of SW heating processes as well as dynamical coupling
      with the troposphere in global climate models is critically important.

\subsection{Impact of  SSI variability in climate models}
\label{sec:42}
      The uncertainty of SSI variations in recent observations and models
      has significant influence on simulations of the climate system, since
      the response of the atmosphere strongly depends on the spectral
      distribution of the solar irradiance. The effects for middle
      atmosphere heating, ozone chemistry and middle atmospheric
      temperatures are examined in the following.

\subsubsection{Effects on atmospheric heating and ozone chemistry }
\begin{figure}
\includegraphics[scale=0.48, angle=270]{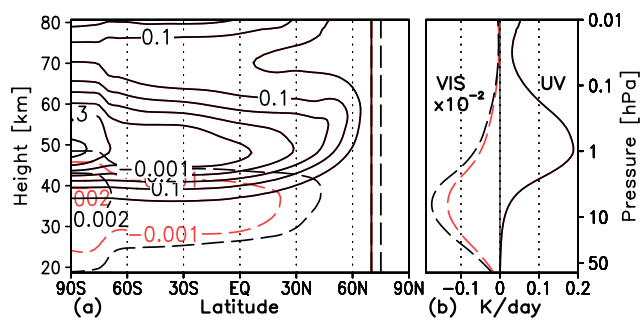}
\caption{Updated from \citet[][]{Oberlaenderetal:2012} \textbf{(a)} Difference in zonal mean solar
 heating rates between 2004 and 2007 in \unit{K\,day^{-1}} from the
SIM data, and \textbf{(b)} difference in the global mean 2004--2007
 solar heating rate signal between { SORCE} and NRLSSI data.
 Solid lines denote the UV spectral range, and dashed lines denote the VIS spectral range.
 Red dashed lines in panels \textbf{(a)} and \textbf{(b)} denote the same differences in zonal and global mean,
 as calculated and presented in Figs.~3b,~c in \citet[][]{Oberlaenderetal:2012};
black dashed lines indicate the enhanced solar signal in the VIS range, when extending the spectral resolution of the FUBRad shortwave
radiation code from one to six bands in the Chappuis bands. { The simulations were performed for perpetual January conditions.} }
\label{fig:tosca_fig4_1}
\end{figure}

      As solar radiation is the primary source of energy that drives
      atmospheric as well as oceanic circulation, accurate representation of
      solar irradiance is of paramount importance for the simulation of the
      atmospheric temperature, composition and dynamics in climate models.
      The variability of the solar spectrum in time, and in particular over
      the solar cycle is necessary for the assessment of solar influence on
      climate. The amplitude of the simulated solar signal depends on the
      spectral solar fluxes prescribed at the TOA. Differences in the TOA
      solar irradiance spectrum result in large changes in the heating rates
      calculated by SW radiation schemes or radiative transfer models, as
      has been shown by \citet[][]{Zhongetal:2008} with differences of up to
      $\approx$\,1.1\,\unit{K\,day^{-1}} in mid-latitude summer. Recently,
      \citet[][]{ Oberlaenderetal:2012} examined the impact of a~number of
      different estimates of prescribed TOA solar fluxes on the solar
      response in a~GCM which includes a~radiation scheme with enhanced
      spectral resolution \citep[][]{Nissenetal:2007} and is therefore able
      to accurately represent the solar signal induced changes. They used
      the NRLSSI, the SATIRE and the SCIAMACHY solar flux input data sets,
      and compared their effects on SW heating rates over the 11-yr solar
      cycle using offline calculations.  They also calculated the
      corresponding temperature response from perpetual January GCM
      simulations with prescribed ozone concentrations.

      The comparison revealed clear differences in SW heating rates for the
      solar minimum of cycle 22 (September 1986). The simulations forced
      with the NRLSSI reconstructions show the smallest solar heating
      rates. The use of the SATIRE reconstructions leads to stronger solar
      heating of up to 5\,{\%} in the middle and upper stratosphere. The
      SCIAMACHY observations slightly enhance the solar heating in the
      mesosphere, with differences arising from the stronger solar fluxes in
      the Huggins bands of the SATIRE model and enhanced fluxes in the
      Hartley bands of the SCIAMACHY data set. Using   SORCE (SIM and SOLSTICE for wavelengths below 210 nm)  measurements
      over the period May 2004 to November 2007 reveals larger changes in
      solar heating rates and the resulting temperatures in comparison to
      the NRLSSI data. The lower
      irradiance in the visible range at higher solar activity than at
      minimum activity in the SORCE/SIM data does not lead to a~decrease in
      total radiative heating.

      The spectral resolution of the SW radiation scheme of EMAC-FUB has
      recently been extended towards a~more detailed representation of the
      Chappuis bands, and an  update of
      \citet[][]{Oberlaenderetal:2012}, performed with the extended spectral
      resolution for solar minimum conditions in November 2007, separated
      for the UV and VIS spectral changes is presented in
      Fig.~\ref{fig:tosca_fig4_1}. The enhanced UV irradiance in 2004
      (compared to 2007) in the { SORCE}  data leads to higher SW heating
      rates while lower heating rates are simulated in the visible spectral
      range (Fig.~\ref{fig:tosca_fig4_1}a). { The SW heating rate change from
      2004 to 2007 is stronger when the model is forced with the SORCE than with the NRLSSI data }
      by 0.18\,\unit{K\,day^{-1}} in the global mean
      (Fig.~\ref{fig:tosca_fig4_1}b).  An enhanced sensitivity to changes in
      the Chappuis bands is found, a~result which illustrates better the
      heating rate changes in the UV, visible and NIR spectral regions from using the { SORCE} data.

  The different magnitude and
      spectral composition of the SSI changes leads not only to
      a~substantial alteration of heating rates considered above but also
      affects photolysis rates which drive atmospheric chemistry and
      regulate the atmospheric ozone distribution in the middle atmosphere
      \citep[e.g.][]{BrasseurSolomon:2005}. The global ozone abundance is
      maintained by ozone production, destruction and transport by air
      motions. However, in the tropical stratosphere above
      $\sim$\,30\,\unit{km} the ozone concentration depends primarily on
      photochemical processes, with oxygen photolysis playing a~crucial role
      in atmospheric chemistry. Therefore, the spectral composition and the
      magnitude of the SSI changes are of critical importance, since they
      define not only the magnitude but also the sign of the direct ozone
      response \citep[e.g.][]{Rozanovetal:2002}.  The solar induced net
      effect of ozone in the stratosphere depends on the competition between
      ozone production due to oxygen photolysis in the Herzberg continuum
      (185--242\,\unit{nm}) and ozone destruction caused by the ozone
      photolysis in the Hartley band (between 200 and 300\,\unit{nm} in the
      UV).

\subsubsection{Effects on ozone and temperature from atmospheric model simulations }
\begin{table*}[t]
\caption{Overview of atmospheric chemistry-climate models and the respective experimental designs using the NRLSSI and/or the SORCE data to study the atmospheric impact.}
\label{table41}
\scalebox{.7}[.7]{
\begin{tabular}{p{30mm}p{20mm}p{20mm}p{18mm}p{24mm}p{14mm}p{26mm}p{20mm}p{38mm}}
\tophline
{Model}  &{Horizontal resolution} & {Top level} & {Number   of    layers} & {Ozone   interactively   calculated} & {QBO} & {Length of   simulation} & {Equilibrium   simulation} & {Reference}\\
\middlehline
\multicolumn{9}{c}{{SORCE forcing comparison}}\\[6pt]
EMAC-CCM    with FUBRad &T42 & 0.01\,\unit{hPa} & 39 & no & no & 50\,\unit{yr} & perpetual Jan & \citet[][]{Oberlaenderetal:2012}\\
GEOS CCM   (using GEOS5) & 2.5{\degree}\,$\times$\,2.0{\degree} & 0.01\,\unit{hPa} & {72} & {yes} & {no} & {25\,\unit{yr}} & {yes} & \citet[][]{Swartzetal:2012}\\
HadGEM3 & {1.875{\degree}\,$\times$\,1.25{\degree}} &{85\,\unit{km}} &{85} & {no} & {yes} & {20\,$\times$\,4\,\unit{yr} ens} & {yes} &  \citet[][]{Inesonetal:2011}\\
IC2-D &  {9.5{\degree}\,lat} & {0.26\,\unit{hPa} for   chemistry} & {17 for   chemistry} & {yes} &{no} & {670\,\unit{days}} & {yes} & \citet[][]{Haighetal:2010}\\
SOCOL v2.0 &  {T30} & {0.01\,\unit{hPa}} & {39} &{yes} & {nudged} & {62\,\unit{months}} &{transient run (01.2004--02.2009)} &  \citet[][]{Schraneretal:2008,Shapiroetal:2012c}\\
WACCM & {2.5\textdegree{} x 1.89\textdegree} & {0.000006 hPa} & {66} & {yes} & {yes} & {25 years} &{yes} &  \citet[][]{Merkeletal:2011}\\[12pt]
\multicolumn{9}{c}{{CMIP5 simulations}}\\[6pt]
HadGEM2 & 1.875{\degree}\,$\times$\,1.25{\degree} & 85\,\unit{Km} & 60 & prescribed solar induced ozone variation & yes & 3 historical ens. (1850--2005) & no & \citet[][]{Hardimanetal:2012}\\
MPI-ESM-LR   (atmospheric   component:   ECHAM6) &T63 & 0.01\,\unit{hPa} & 47 & prescribed   solar induced   ozone variation  &no & 3 historical ens. (1850--2005) & no & H.~Schmidt   (personal communication, 2012) \\
WACCM$^{\ast}$ CESM1 (WACCM) & 2.5{\degree}\,$\times$\,1.89{\degree}{} &  0.000006\,\unit{hPa} & 66 & yes & nudged & 3 historical ens. (1960--2005)& no & \citet[][]{Marshetal:2012}\\
\bottomhline
\end{tabular}
}
\end{table*}

\begin{table*}[t]
\caption{ Overview of the spectral range of forcing for the various atmospheric chemistry-climate models employed to study the atmospheric impact.}
\label{table42}
\scalebox{.9}[.9]{
\begin{tabular}{p{85mm}p{100mm}}
\tophline
{Model}  &{Spectral range of forcing}\\
\middlehline
\multicolumn{2}{c}{{SORCE forcing comparison}}\\[6pt]
EMAC-CCM with FUBRad & 1) NRLSSI\\
&  2) SIM (SOLSTICE $<$210\,\unit{nm})\\
GEOSCCM (using GEOS5) &  1) NRLSSI\\
 &  2) SOLSTICE $<$244\,\unit{nm} and SIM 244 to 1600\,\unit{nm}\\
 &  Solar max and solar min conditions inferred by linear fitting approach based on the available SORCE time series. \\
HadGEM3 &  SIM (only 200-320\,\unit{nm})\\
& Extrapolated  in time to represent the full solar-cycle amplitude, The estimated 4\% change is evenly distributed in the UV band. No changes in other bands. \\
IC2$-$D &  1) NRLSSI\\
 &  2) SIM (SOLSTICE $<$200\,\unit{nm}) \\
SOCOL v2.0 &   1) NRLSSI\\
 &  2) SOCOL SIM (SOLSTICE $<$210\,\unit{nm} and SIM 210-750\,\unit{nm})\\
 &  3) SOCOL SOLSTICE (SOLSTICE $<$ 290\,\unit{nm} and SIM 290-750\,\unit{nm})\\
WACCM &  1) NRLSSI\\
&  2) SOLSTICE (116-240\,\unit{nm}), SIM (240-2000\,\unit{nm}) and SRPM (2000-11000\,\unit{nm})\\[12pt]
  \multicolumn{2}{c}{CMIP5 simulations}\\[6pt]
HadGEM2 &  NRLSSI (CMIP recommended) \\
MPI-ESM-LR (atmospheric component: ECHAM6) &  NRLSSI (CMIP recommended)  \\ 
WACCM$^{\ast}$ CESM1 (WACCM) &  NRLSSI (CMIP recommended)\\
\bottomhline
\end{tabular}
}
\end{table*}

\begin{figure*}
\includegraphics[width=85mm]{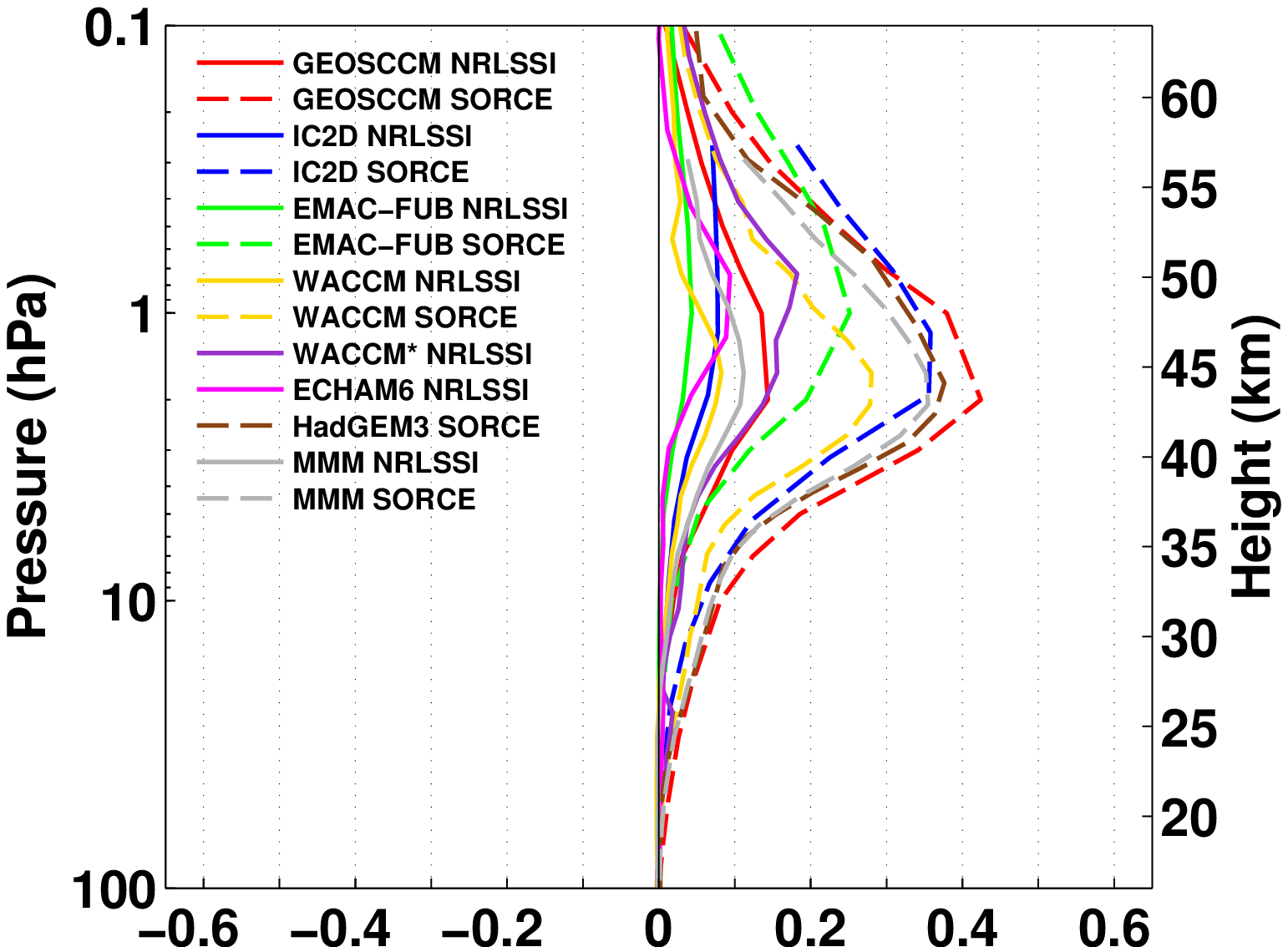}
\includegraphics[width=85mm]{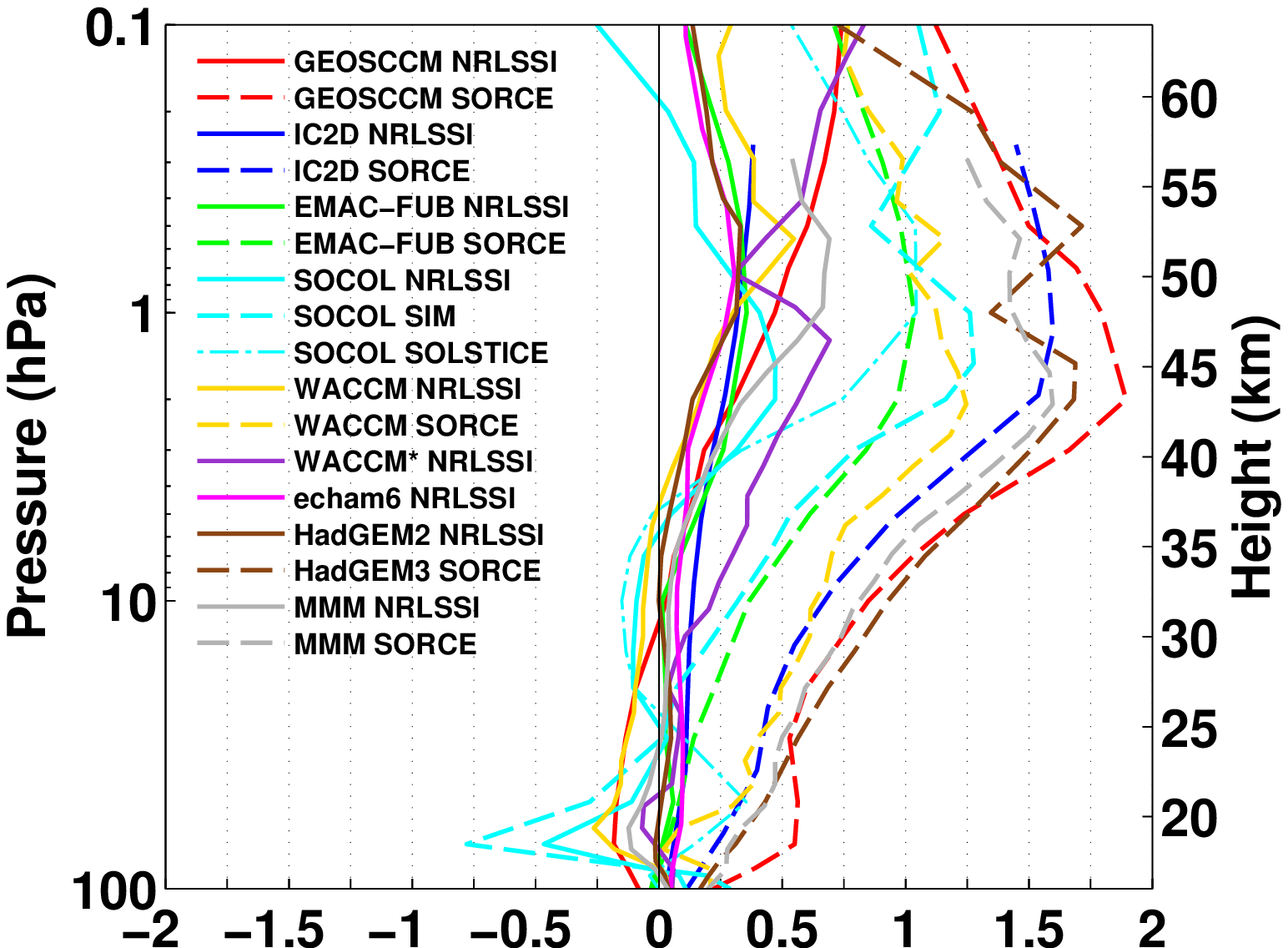}
\caption{Differences for January between 2004 and 2007
 averaged over the tropics from 25{\degree}\,S to 25{\degree}\,N
for different atmospheric models using the NRLSSI data (solid lines)
 and the SORCE/SIM and SORCE/SOLSTICE data (dashed lines) for \textbf{(a)}  the shortwave heating
rate in Kelvin per day (\unit{K\,day^{-1}}) and \textbf{(b)}~the temperature in Kelvin (K). For details please see text.}
\label{fig:tosca_fig4_2}
\end{figure*}

\begin{figure}
\includegraphics[width=85mm]{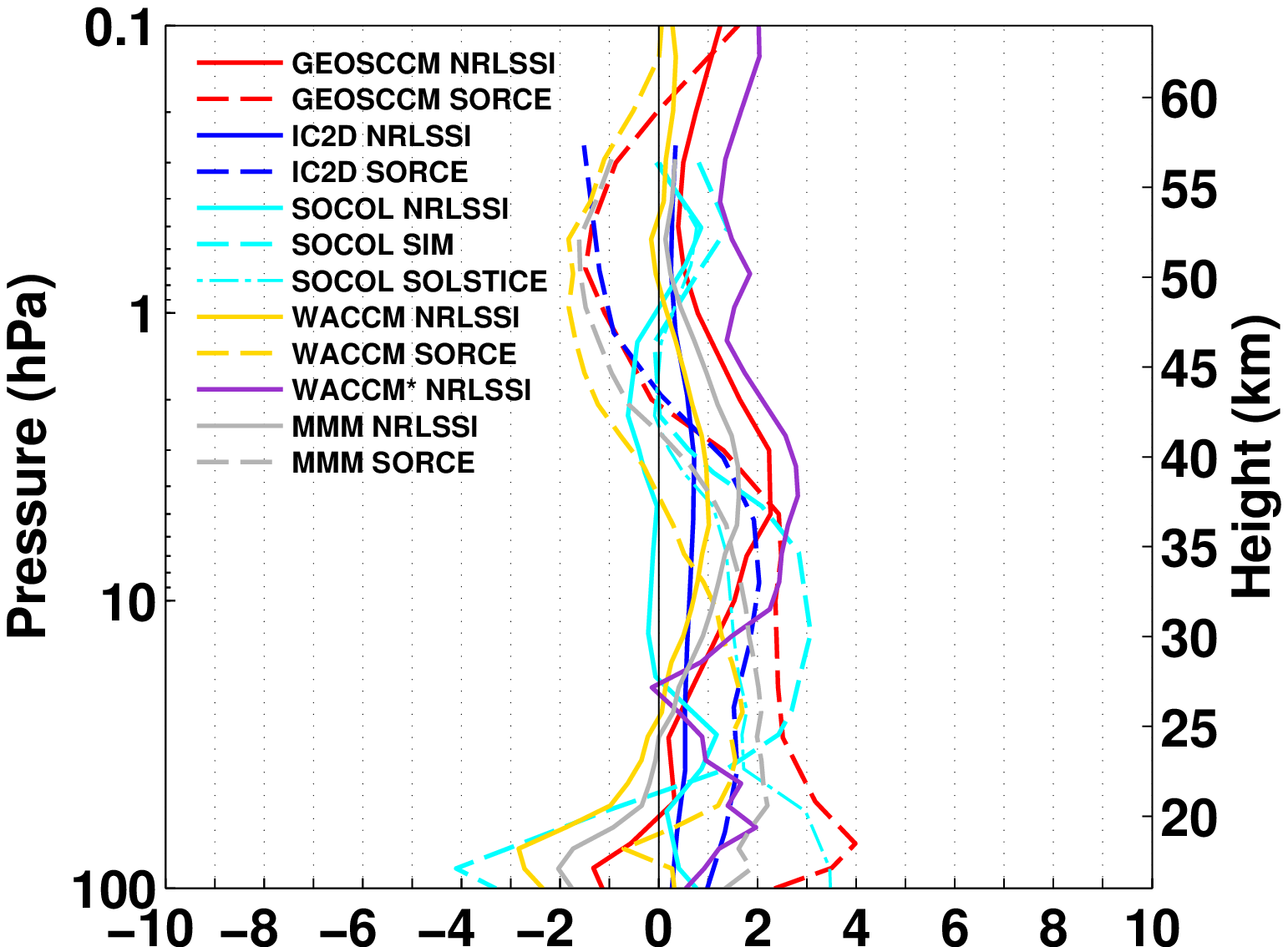}
\caption{As Fig.~\ref{fig:tosca_fig4_2} but for the ozone response in percent.}
\label{fig:tosca_fig4_3}
\end{figure}
      
      The effects of different SSI estimates on stratospheric ozone
      as deduced by CCMs or 2-D radiative-photochemical model simulations
      have been recently reported in a~number of papers which are discussed and compared to each other in the following. In particular, a
      ~number of analyses  focused on the differences in atmospheric response
      between simulations using the SORCE  and the NRLSSI model results. The
      NRLSSI is the data set widely used for climate modelling purposes,
      including the SPARC CCMVal (Chemistry-Climate Model Validation) and
      the CMIP5 simulations for the upcoming IPCC
      report \citep[][]{Tayloretal:2012}.  \citet[][]{Cahalanetal:2010} and \citet[][]{Haighetal:2010}
 published first the
      important implications of the SORCE data for middle atmosphere
      heating and therefore temperatures. Using simulations  from a~2-D radiative-photochemical
      model \citet[][]{Haighetal:2010} also presented effects on ozone. A~number of other studies using 3-D chemistry-climate models
      followed \citep[][]{Merkeletal:2011,Inesonetal:2011,
      Matthes:2011,Oberlaenderetal:2012,Swartzetal:2012}.  To better
      understand the atmospheric sensitivity to the range of SSI estimates
      discussed above, these recently published model experiments using the
      NRLSSI and the SORCE data are compared to each other in
      Figs.~\ref{fig:tosca_fig4_2} and~\ref{fig:tosca_fig4_3} with respect
      to the solar signal in SW heating rates, temperatures and ozone.
      Additionally,  results from three CMIP5 model experiments
 using the NRLSSI data set are shown.  Table \ref{table41} presents an overview of the models and their experimental designs. Details on the spectral solar forcing used in each simulation are given in Table  \ref{table42}.
 The purpose is to provide the reader with an initial comparison of the impacts of two SSI data sets which represent  the
 lower and upper boundary of SSI variations. One has to keep in mind, though, that all models used slightly different 
 experimental setups { (including slightly different spectral ranges for the SORCE data) and therefore an exact comparison between them as well as to observations awaits common coordinated experiments. Moreover, we note that the solar spectral forcing from the SORCE satellite used by all the models is derived from a relatively short period, and longer model simulations with full solar cycle forcing would be needed in order to better evaluate the differences in the response. } \\

\noindent
\textbf{Response in shortwave heating rates}\\

\noindent
      Figure~\ref{fig:tosca_fig4_2} shows the differences in SW heating
      rates (top) and temperatures (bottom panel) for the month of January
      in the tropical region, i.e. averaged between 25{\degree}\,S and
      25{\degree}\,N. Differences are calculated between January of the year
      2004 (during the declining phase of solar cycle 23), and January of
      2007, close to solar minimum in December 2008, using the NRLSSI data
      (solid lines) and the SORCE/SIM and SORCE/SOLSTICE data (dashed
      lines).

  Most of the models listed in the upper part of Tables \ref{table41} and \ref{table42} performed equilibrium simulations 
for the years 2004 and 2007. 
EMAC performed perpetual January runs only, while 
SOCOL performed transient simulations from 2004 to 2009, with the differences shown here  calculated between January of the year 2004 and January of 2007. Therefore most of the displayed differences
 do not provide one full solar cycle signal. The HadGEM3 model simulation used the SORCE/SIM measurements and 
 scaled the data from 2004 to 2007 to obtain a full solar cycle signal \citep[][]{Inesonetal:2011}.  { GEOSCCM simulations were  performed for solar maximum and minimum conditions as well, inferred by a linear fitting approach based on the available SORCE time series \citep[][]{Swartzetal:2012}. The  differences presented for HadGEM3 and GEOSCCM} are
 slightly larger than for the other models { (and closer to each other), because the simulations correspond to}   the full solar cycle signal.  Please note that SOCOL simulations were performed with the NRLSSI as well as combinations of SORCE/SIM and SORCE/SOLSTICE data \citep[][]{Shapiroetal:2012c}. 
 The 
 simulations performed with three CMIP5 models listed in the lower part of  Tables \ref{table41} and \ref{table42} represent ensemble mean time series experiments with varying 
 solar cycle \citep[e.g.][]{Hardimanetal:2012}.
In order to be comparable to the
      other equilibrium simulations, the differences shown represent the
      solar signal from a~multiple linear regression analysis using the
      monthly mean TSI or Hartley band irradiance as a~solar
      proxy. Afterwards the resulting solar signal has been scaled to the
      2004 to 2007 value, i.e. 24\,{\%} of the full solar cycle. Note that
      not all the models were able to provide SW heating rates and/or used
      interactively calculated ozone. Therefore only a~model subset is
      displayed in the respective figures. In addition to the lines for the
      single model realisations a~multi model mean (MMM) for three models,
      namely WACCM, GEOSCCM, and HadGEM3 is included in
      Figs.~\ref{fig:tosca_fig4_2} and~\ref{fig:tosca_fig4_3}.  As expected,
      the largest SW heating rate response occurs near the stratopause
      ($\sim$\,1\,\unit{hPa}) showing the direct impact of solar irradiance
      variations from 2004 to 2007.  Although the strength of the response
      differs between the different models, all models show a~consistently
      stronger SW heating (by a~factor of up to two to three) using the
      SORCE data as compared to the NRLSSI data. For both NRLSSI and
      SORCE simulations, the GEOSCCM shows the strongest response with
      SW heating rate differences of up to 0.12\,\unit{K\,day^{-1}}
       for NRLSSI compared to
      0.42\,\unit{K\,day^{-1}} for SORCE closer to the HadGEM3 response, as both  models are depicting full solar cycle conditions. In contrast to the other models,  HadGEM3, the EMAC-FUB did not
      take into account solar induced ozone variations. EMAC therefore shows the weakest  SW response in both experiments, followed by the WACCM
      model. The simulations performed with the CMIP5 models fit nicely to
      the equilibrium simulations with the same  (NRLSSI) forcing. Given the fact that
      the NRLSSI data provide a~lower boundary for SSI variations with the
      solar cycle, the SW heating rate effects are likely enhanced in nature
      as compared to recent chemistry-climate model simulations. The SW  heating rate
      differences described above may be attributed to the difference in the
      radiation codes of the models, as discussed    in the previous section, but are also  due to the different
 model setups  (for example, in simulations performed with or without interactive ozone, the signal will be enhanced or diminished). \\ 

\noindent 
\textbf{Response in temperatures}\\

 \noindent
 Corresponding to the
      SW heating rate differences, the tropical temperature differences for
      January are largest near the stratopause; using  SORCE, the
      temperature differences between 2004 and 2007 are three to four times
      larger than with NRLSSI. Again, the GEOSCCM and the HadGEM models
      produce the strongest response with about 2\,\unit{K} and the WACCM
      and the FUB-EMAC model the weakest response. The response in SOCOL
      differs slightly from that of the other models by showing the maximum
      effect around 0.3\,\unit{hPa}, i.e. at a~higher altitude than the
      other model simulations. In addition, the temperature response in
      SOCOL is always weaker using SORCE/SIM or NRLSSI data. 
 When using SORCE/SOLSTICE spectral forcing the temperature response is approximately half of the response obtained using SORCE/SIM, indicating a large sensitivity of the model to the SSI data sets  depicting their differences. Although none of the other models has performed a similar set of simulations, it is expected that this sensitivity is typical for all models shown here.  Please note that the negative temperature response in SOCOL in the lower and middle stratosphere is not statistically significant. The temperature responses in the CMIP5 models closely agree with the NRLSSI equilibrium simulations. { Comparison of the temperature response to observations over a full 11-yr solar cycle \citep[e.g.][]{Austinetal:2008,Remsberg:2008a} reveals that the signals derived with the use the NRLSSI data generally shows a response peaking around the stratopause, with altitudes of maximum response varying between models according to their model setups as well as the use of solar induced ozone variations. As presented in Chapter 8 of the SPARC-CCMVal Report (2011), where the solar cycle effect on stratospheric temperature from a number of CCM simulations is discussed, the response to the NRLSSI forcing over the full solar cycle is about 0.8K, similar to observations. The response to the SORCE data forcing is larger than in the observations, but does not scale linearly with larger forcing. It was found to be higher by almost a factor of two in the case of the GEOSCCM simulation which corresponds to a full solar cycle \citep[][]{Swartzetal:2012}. \\

\noindent
\textbf{Response in ozone}\\  

\noindent
      Figure~\ref{fig:tosca_fig4_3} displays the tropical ozone response in
      percent for month of January for those models that calculate ozone
      interactively. In the lower stratosphere,
      ozone changes induced by the forcing with { SORCE} data are larger
      and remain positive as compared to those calculated with NRLSSI
      data. In the middle to upper stratosphere ozone differences induced by
      the NRLSSI data are smaller but still positive whereas the ozone
      response using the { SORCE} data is negative. The height of the
      change from positive to negative ozone response varies from
      5\,\unit{hPa} to 2\,\unit{hPa} between the models. Observations from
      8\,\unit{yr} of SABER data indicate a~transition altitude of about
      1\,\unit{hPa} \citep[][]{Merkeletal:2011} that is higher than for all
      models. This change of sign in the ozone response using SORCE data,
      meaning lower ozone during solar maximum than during solar minimum in
      the upper stratosphere, is statistically significant in all
      models. \citet[][]{Swartzetal:2012}, who also examined the total ozone
      response and contributions from heating and photolysis, show good
      agreement between observations and simulations with the GEOSCCM
      model. \citet[][]{Haighetal:2010} showed that the ozone decrease in the
      upper stratosphere and mesosphere is related to photochemical
      processes. Decreased \chem{O_3} at solar maximum is consistent with
      increased \chem{HO_x} and O and also leads to a~self-healing effect
      with more UV radiation reaching lower levels, enhancing \chem{O_2}
      photolysis and therefore increased \chem{O_3}. Additional
      observational evidence is needed to confirm the sign reversal in upper
      atmospheric ozone response.\\

\noindent
\textbf{Dynamical changes and impact on the troposphere}\\
\begin{figure*}
\includegraphics[width=85mm]{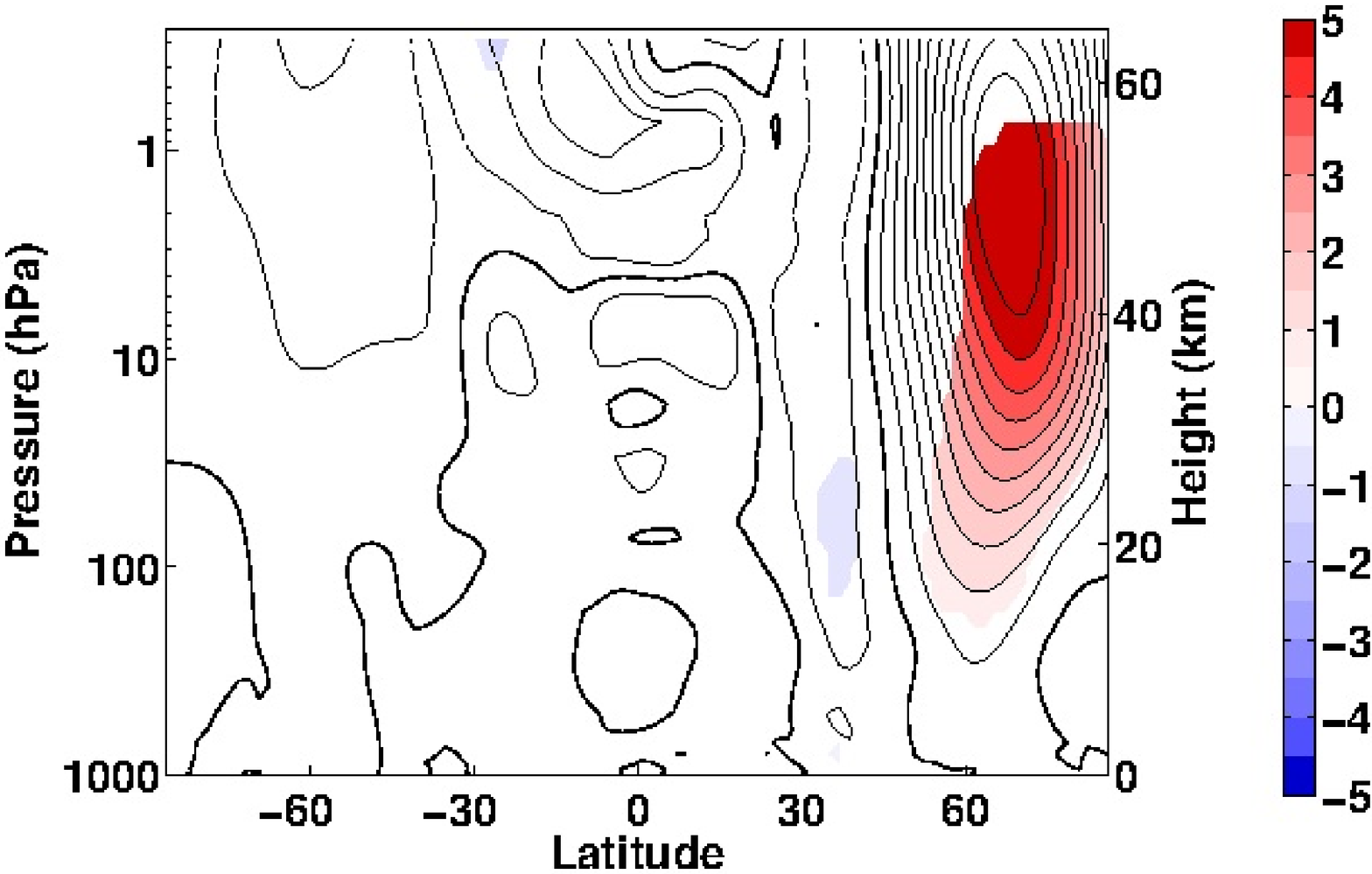}
\includegraphics[width=85mm]{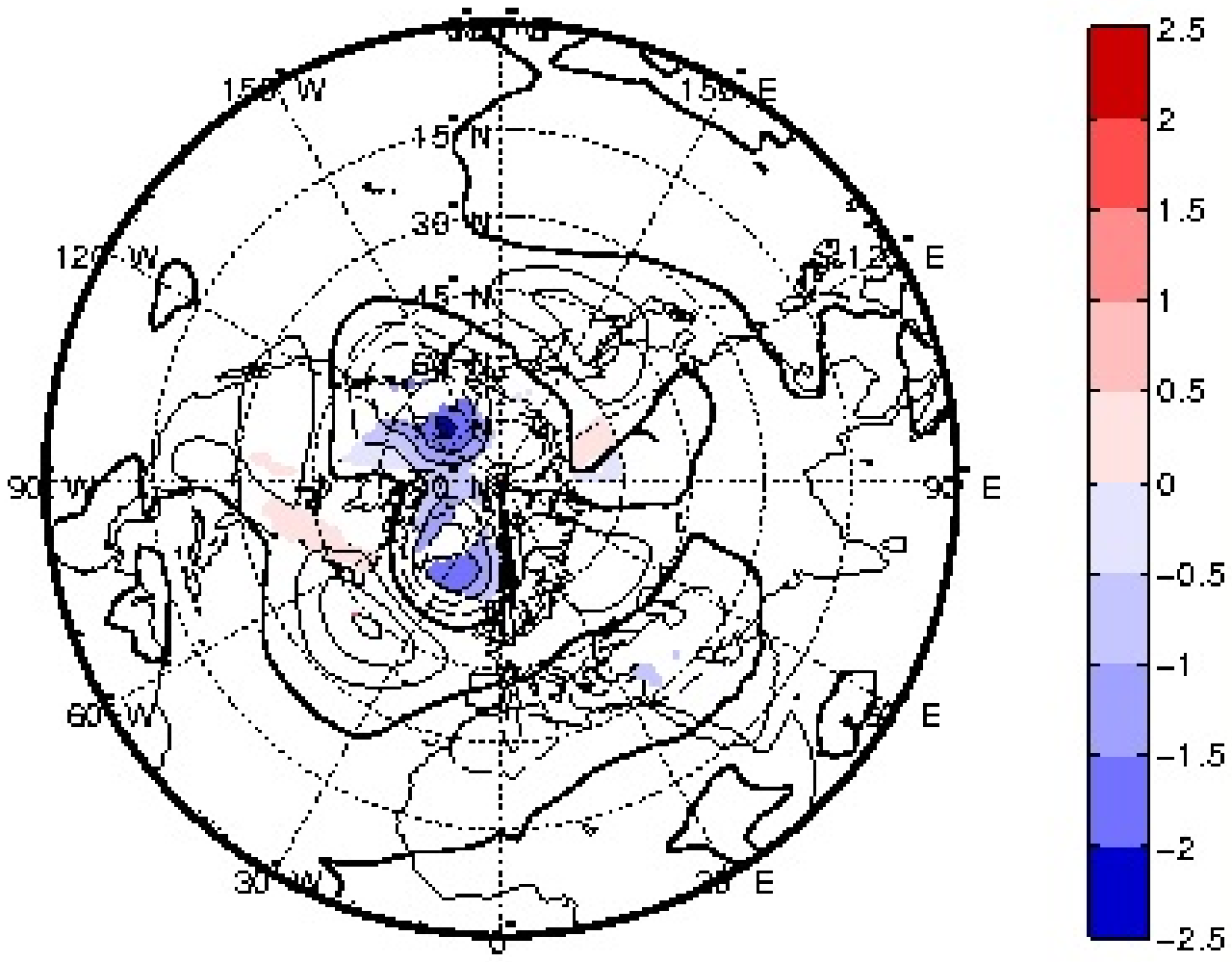}
\caption{Multi Model Mean (MMM) of the HadGEM3, the GEOS CCM, and the WACCM model for \textbf{(a)}  the zonal mean zonal wind differences between 2004 and 2007 in January using the { SORCE} forcing, and \textbf{(b)}~the corresponding geopotential height differences at 500\,\unit{hPa}. { Coloured areas provide 95\% statistically significant signals.} }
\label{fig:tosca_fig4_4}
\end{figure*}

\noindent
      So far the discussion focused on the response of stratospheric
      heating, temperatures and ozone to the forcing with NRLSSI and
     SORCE data.  The direct solar response in the upper stratosphere
      induces indirect circulation changes throughout the stratosphere
      \citep[e.g.][]{Kuroda:2002} and also affects the troposphere and the
      surface \citep[][]{Haigh:1999, Kodera:2002, Matthesetal:2006}.  The
      impact on the troposphere using { SORCE} data has been reported by
      \citet[][]{Inesonetal:2011}, who provide intriguing but provisional
      results on northern hemispheric winter
      circulation. Figure~\ref{fig:tosca_fig4_4} displays January
      multi-model mean signals for three climate models (GEOSCCM, HadGEM3,
      and WACCM) in zonal mean zonal wind as well as 500\,\unit{hPa}
      geopotential height using { SORCE data} as external
      forcing. A~stronger  and statistically significant  polar night jet during solar maximum years
      dominates the stratosphere and reaches down to the troposphere,
      whereas negative wind anomalies { that are statistically significant in the lower stratosphere and upper troposphere only} dominate equatorwards
      (Fig.~\ref{fig:tosca_fig4_4}, top panel).  The zonal wind signals
      correspond to a~positive Arctic oscillation (AO) pattern with
      a~{ statistically significant} stronger polar vortex and  a partly statistically significant  enhanced geopotential heights in
      midlatitudes (Fig.~\ref{fig:tosca_fig4_4}, bottom panel) in agreement
      with observations \citep[e.g.][]{Kodera:2002}. The stronger polar
      vortex in the stratosphere leads to a~stronger positive phase of AO
      and North Atlantic Oscillation (NAO), which means a~stronger Iceland
      low and higher pressure of the Azores and hence an amplified storm
      track over Europe. This in turn leads to mild conditions over northern
      Europe and the eastern US and dry conditions in the
      Mediterranean. However, the tropospheric AO and NAO already respond to
      lower UV irradiance variations over the 11-yr cycle, as is shown in
      \citet[][]{Langematz:2012} who obtained a~similar AO/NAO response from
      a~transient EMAC-FUB simulation for the period 1960--2005 using NRLSSI
      data as TOA input.  The signal is seen also in
      \citet[][]{Matthesetal:2006}, but in equilibrium simulations with
      NRLSSI as TOA input. As shown by \citet[][]{Inesonetal:2011} these
      patterns imply that the solar cycle effect on the AO/NAO contributes
      to a~substantial fraction of the typical year-to-year variations and
      provides therefore a~potentially useful source of improved decadal
      climate predictability for the Northern Hemisphere. Note that the
      response is regional and is negligable on the global average. However,
      a~caveat is that the 11-yr solar cycle variability cannot be
      forecasted into the future (on a~daily, yearly or decadal time scale).

      Even though the solar variability on time scales longer than the 11-yr
      solar cycle is beyond the scope of this paper, we should note here
      again that the Sun is the fundamental energy source of the climate
      system. As such, the low solar activity in the past few years
      (compared to the previous 6 solar cycles), and its possible
      implications for future climate evolution has attracted the attention
      of both scientists and the public \citep[e.g.][]{Lockwoodetal:2010,
      Schrijveretal:2011, Jonesetal:2012, Rozanovetal:2012}.

\subsection{Discussion of CCM results}
\label{sec:43}

      We  described   the impact of NRLSSI and
      SORCE data, which represent the lower and upper boundaries of SSI solar cycle
      estimated variations, on the atmosphere and climate as depicted in
      CCM  simulations. The NRLSSI reconstructions
      provide the standard data base for simulations of the recent past and
      future \citep[e.g.][]{SPARC-CCMVal:2010,Tayloretal:2012}.  The
      atmospheric response with respect to this standard data set is
      compared to    that derived from a different SSI estimate  to understand not only
      the single model responses, but  also to point out   the
      importance and robustness of solar cycle signals for climate
      simulations.
  In particular, it is worth mentioning that an  enhanced
      spectral resolution in the radiation codes leads to enhanced
      sensitivity in the response.   We also described  the important role of solar
      induced ozone changes for the amplification of solar effects on
      atmospheric composition, circulation and climate.
      
      Model simulations using the  SORCE measurements as compared to the
      NRLSSI model show larger (by a~factor of~two) SW heating and temperature
      signals. The lower irradiance in the visible range during higher solar
      activity than at minimum activity in the  SORCE  measurements does
      not affect the increase in total radiative heating. Recent atmospheric
      model simulations with enhanced spectral resolution, however, point to
      the importance of the Chappuis bands in lower stratospheric
      heating. The solar ozone signal derived when the NRLSSI data are used
      is positive throughout the stratosphere and mesosphere, whereas the
      sign is reversed in the upper stratosphere when the { SORCE}  data are
      used.  Observations  indicate a~reversal of sign in the solar cycle ozone response
      in the upper atmosphere, however, the transition altitude for the sign
      change is higher than models suggest. This result needs to be
      confirmed by other satellite ozone measurements as observational evidence of the solar ozone signal is still limited.  Finally, the
      tropospheric and surface solar cycle response has been presented in an
      ensemble mean and it has been highlighted that while these solar induced changes are
      of minor importance for globally averaged temperatures there are
      larger regional responses.

\conclusions
\label{sec:concl}

This paper presents an overview of our present knowledge of the impact
of solar radiative forcing on the Earth's atmosphere. It covers the
observations and the modelling of the solar radiative input as well as
the modelling of the Earth's atmospheric response. The focus is on
satellite-era (i.e. post-1970) data for which direct solar irradiance
observations are routinely available. Special attention is given to
the role of the UV spectral region, whose small contribution to TSI is
compensated by a~high relative variability with a~potentially
amplified influence on climate through radiative heating and ozone
photochemistry.
There is today clear evidence for the impact of solar
variability on climate but both its magnitude and its confidence level
are still subject to considerable debate. One major challenge lies in
the extraction of the weak solar signal from the highly variable
atmospheric state. Recent progress has been made along two
directions. The first one is the assessment of the magnitude of
secular trends in solar radiative forcing through the reconstruction
of solar activity on centennial and millenial time scales from
indirect indices such as cosmogenic isotopes,
\citep[e.g.][]{Solankietal:2004,BardandFrank:2006,Beeratal:2006,Usoskin:2008,Schmidtetal:2012}. The
second one, which we focus on, deals with shorter time scales, and is
about recent solar variability and its impact on the lower and middle
atmosphere.
The global physical mechanisms that cause the solar irradiance to
change in time and eventually impact climate have been well
documented, \citep[e.g.][and references
therein]{Haighetal:2005,Haigh:2007,Grayetal:2010,LeanandWoods:2010}. More
than three decades of SSI observations are now available, but they are
highly fragmented and agree poorly because of the difficulties in
making radiometrically calibrated and stable measurements from
space. Merging these different observations into one single and
homogeneous record is a~major and ongoing effort. As a~consequence,
several models have been developed for reproducing the SSI and its
variability. The most successful ones are semi-empirical models such as SATIRE, SRPM, and COSI, which 
describe the SSI in terms of contributions coming from different solar
surface magnetic features such as sunspots and faculae. Most models
nowadays reproduce SSI measurements on short term time scales fairly
well. However, uncertainties in SSI changes still remain on long term
time scales and in the 220--400\,\unit{nm} band, which is of
particular interest because of its impact on stratospheric ozone.
These modelled or observed variations in the SSI are today used as
inputs to CCM simulations that are capable of properly reproducing
most aspects of stratospheric heating and point to the existence of
a~significant impact of solar variability on climate. However, major
uncertainties remain in their detailed description, in which nonlinear
couplings and regional effects can play an important role.

The main topics discussed in this paper and main conclusions of our study are:
\begin{itemize}
\item The spectral and temporal coverage of the SSI measurements gradually improved over time, and with the full operation of the SORCE mission in April 
2004,  daily observations of the full UV, visible and NIR
spectrum became available. Unfortunately, this situation is likely to end in
2013, when SORCE is anticipated to succumb to battery failure. The lack of  SSI observations has led to
intensive application of semi-empirical models. There remains
a~considerable issue in assimilating SSI observations in such models and in
reconstructing the SSI prior to the space age.  
Reconstructions going back to
the early 20th century  can be derived from  semi-empirical  models  based on historical ground-based
solar observations \citep[e.g.][]{Ermollietal:2009}.

\item Recent SSI measurements by SORCE  suggest a larger (factor 2-6) variability in the UV (200-400~nm) during solar cycle 23, which is hard to reconcile with earlier
SSI and TSI observations (e.g. by UARS/SOLSTICE and UARS/SUSIM) and  with SSI models. However, new estimates of possible calibration corrections suggest that the UV irradiance variation derived from SORCE measurements might be almost a factor of two weaker than reported earlier \citep[][]{Woods:2012}.

\item TSI alone does not adequately describe the solar forcing on the
atmosphere and therefore SSI variations have to be taken into account in climate models. For many years, the canonical value of the
average TSI was 1365.4\,$\pm$\,1.3\,\unit{W\,m^{-2}} whereas now the most
accurate, and generally accepted, value is 1361\,$\pm$\,0.5\,\unit{W\,m^{-2}}
\citep[][]{KoppLean:2011,Schmutzetal:2012}.

\item There has been steady progress in modelling SSI
variations, and a number of models are now available that can be used as
input for climate studies. However, the main uncertainty in the models concerns the wavelength range 220--400~nm, where the magnitude of the variations differs by as much as a factor of three between models. This range is of particular interest for climate studies. 
Of the five SSI models discussed, specifically NRLSSI, SATIRE-S, COSI,
SRPM and OAR, only one (SRPM) shows a behaviour of the UV and visible irradiance
qualitatively resembling that of the recent SORCE/SIM measurements. However, it
should be noted that the integral of the SSI computed with SRPM over the entire spectral
range (i.e. the TSI) does not reproduce the measured cyclical TSI changes. None of the other four models, which are in closer agreement with
each other, reproduces the peculiar behaviour of the UV and visible irradiance observed
by SORCE/SIM. 

\item While there has been major recent progress in better reproducing the SSI changes on short term time scales, there remains now an
important issue in the derivation of realistic confidence intervals, both for the observations and for the model results. Further common metrics should be used for comparing them. 

\item Within the range of recent SSI values from observations and
semi-empirical models, the NRLSSI model and { SORCE} observations  represent,
respectively  the lower and  upper limits in the
magnitude of the solar cycle variation in the UV.

\item Results obtained with CCMs show that the observed increase in UV irradiance at solar activity maximum compared to minimum leads to an increase in atmospheric
heating rates and correspondingly an increase in stratospheric temperatures. This direct atmospheric response is larger for larger UV forcing, i.e. for SORCE as compared to NRLSSI, and is sensitive to resolution in the radiation code. Currently there is insufficient observational evidence to support recent SSI measurements by SORCE on the basis of comparisons between climate model simulations and atmospheric ozone or temperature observations in the stratosphere.
The larger UV forcing also leads to a larger surface response. The surface effect is regional and has little influence on globally averaged temperatures.

\item Accurate representation of the spectral nature of the incoming radiation, especially at wavelengths below 320\,\unit{nm}, and therefore ozone photochemical variations in the model simulations, is important since these changes amplify the atmospheric solar signal. A more accurate representation of the Chappuis absorption band in a radiation scheme revealed enhanced sensitivity in the heating rate response, which again highlighted the importance of a well resolved radiation code.

\end{itemize}

A~unique aspect of this study is the description of the solar
terrestrial connection by an interdisciplinary team of solar and
atmospheric physicists. Progress on this hotly debated issue has often
been hampered by the fact that limitations on observations or on
models are not always properly known outside of a~given scientific
community. For the first time a~comprehensive comparison and
discussion of all relevant SSI measurements and models available for climate studies is presented, as
well as a~first investigation of their impacts on Earth's climate within a number of different CCMs.  
These results highlight the importance of taking into account in future climate studies SSI variations and their effects on the EarthÕs atmosphere.

Major efforts, however, are still needed in each of the three research areas covered in this study. Resolving differences in existing long-term time series of solar measurements is a major challenge; forthcoming initiatives for merging these measurements into single homogenous databases will be of invaluable help (see, e.g. http://projects.pmodwrc.ch/solid/).  Overcoming inaccuracies of current SSI models is another important challenge that will benefit from the renewed interest in the solar radiative output, following the recent  solar minimum. Finally, the realistic evaluation by climate models of the solar impact on the EarthÕs climate is an important issue. For example, we focussed on the effect of the solar forcing without quantifying the impacts on amplification and feedback mechanisms. This should be done in a coordinated set of CCM experiments where  the treatment of SSI inputs to the models are completely specified and results are robustly comparable with each other. Then it will be also possible to investigate the effects of the top-down feedback and for CCMs with an interactive ocean also the bottom-up feedback mechanism.

\begin{acknowledgements}
      This work  is the outcome of a workshop funded by  the  COST Action ES1005 TOSCA
      (\url{http://www.tosca-cost.eu}).  We thank J.~Lean,   J.~Fontenla, and G. Thuillier
      for helpful comments  and kind collaboration. The authors are grateful to W.~Ball,
      S.~Criscuoli, S.~Oberl\"ander and M.~Kunze for providing updated
      results of their work, to F.~Hansen for preparing
      Figs.~\ref{fig:tosca_fig4_2} to~\ref{fig:tosca_fig4_4}, and to
      A.~Kubin, G.~Chiodo, S.~Misios, and A.~Maycock for their active
      participation in the workshop of the COST Action ES1005.  \\
                 We make the use of  SORCE/SIM v17, SORCE/SOLSTICE v12, TIMED/SEE v10, UARS/SOLSTICE v18, UARS/SUSIM v22, SBUV, and SME data from the LISIRD at LASP University of Colorado and members of the instrument teams. Our special
      thanks are to all instrument teams that made all the observations that
      were used in this study, as well as to the modelling groups that
      provided published and partly also unpublished data for this study, in
      particular to J.~Haigh, S.~Ineson, D.~Marsh, A.~Merkel, H.~Schmidt,
      A.~V. Shapiro, and W.~Swartz.   This work is part of the WCRP/SPARC
      SOLARIS initiative. \\
      E.~Rozanov and A.~Shapiro are supported by the
      Swiss National Science Foundation under grant CRSI122-130642
      (FUPSOL). S.~K. Solanki acknowledges support by the Korean Ministry of
      Education, Science and Technology (WCU grant No. R31-10016). The work
      of K.~Matthes is supported within the Helmholtz-University Young
      Investigators Group NATHAN, funded by the Helmholtz Association
      through the President's Initiative and Networking Fund and the
      Helmholtz-Zentrum f\"ur Ozeanforschung Kiel (GEOMAR). The research leading to some results presented here has received funding also from the EC FP7/2007-2013 Program (project 284461, ww.eheroes.eu). \\

        We thank the anonymous reviewers, the editor Dr. W. Ward, and all participants in the discussion forum for their comments that helped improving this paper.
\end{acknowledgements}

\newcommand{\ApJ}{Astrophys.~J.}
\newcommand{\ApJL}{Astrophys. J. Lett.}
\newcommand{\ApJS}{Astrophys. J. Suppl. Ser.}
\newcommand{\AAp}{Astron. Astrophys.}
\newcommand{\AApR}{A\&AR}
\newcommand{\AApSS}{A\&AS}
\newcommand{\ARAA}{Annu. Rev. Astron. Astrophys.}
\newcommand{\ApSS}{Ap\&SS}
\newcommand{\AZh}{AZh}
\newcommand{\Afz}{Afz}
\newcommand{\AJ}{Astron.~J.}
\newcommand{\ApJSS}{ApJS}
\newcommand{\BAAS}{BAAS}
\newcommand{\EPS}{Earth, Planets and Space}
\newcommand{\JGR}{J. Geophys. Res.}
\newcommand{\GRL}{Geophys. Res. Lett.}
\newcommand{\MNRAS}{Mon. Not. R. Astron. Soc.}
\newcommand{\PASJ}{PASJ}
\newcommand{\PASP}{Publ. Astron. Soc. Pac.}
\newcommand{\PSS}{Planet. Space Sci.}
\newcommand{\QJRAS}{Q. J. R. Astron. Soc.}
\newcommand{\SPh}{Solar Phys.}
\newcommand{\CelMech}{Celest. Mech.} 
\bibliographystyle{copernicus}

\newpage\eject

\end{document}